\documentclass{article}

\usepackage[english]{babel}

\usepackage[letterpaper,top=2cm,bottom=2cm,left=3cm,right=3cm,marginparwidth=1.75cm]{geometry}
\usepackage{graphicx}
\usepackage{caption}
\usepackage{amssymb}
\usepackage{amsmath}
\usepackage{amsfonts}
\usepackage{mathrsfs}
\usepackage{color}
\usepackage{ulem}
\usepackage{physics}
\usepackage[english]{babel}
\usepackage[utf8]{inputenc}
\usepackage[T1]{fontenc}
\usepackage{mathptmx}
\usepackage{etoolbox}
\usepackage{bm}
\usepackage{endnotes} 
\usepackage{amsmath}
\usepackage{graphicx}
\usepackage[colorlinks=true, allcolors=blue]{hyperref}

\title{Indefinite causal order with definite Bohmian trajectories for the quantum switch}
\author{Aur\'elien Drezet}
\begin{document}
\maketitle

\begin{abstract}
We analyze the notion of indefinite causal order (ICO) from the standpoint of the deterministic de Broglie--Bohm theory, based on the notion of trajectories in configuration space. More precisely, we study the physics of the quantum switch (QS) interferometer within a Bohmian framework and show how the notion of causality is preserved in this approach. Furthermore, we investigate device-independent certification and causal inequalities applied to an entangled quantum system, and show that in this regime too, the notions of trajectories and of a strict causal order are perfectly preserved.

\end{abstract}

 \section{Introduction}\label{sec:introduction}
 \indent The question of causality---its meaning, its fate---has haunted quantum mechanics since its birth in the 1920s--30s.
Indeed, nearly a century ago, in Brussels in October 1927, the fifth Solvay Conference gathered most of the founders of the new quantum theory \cite{Solvay1928}. Bohr, Heisenberg, Born and Pauli championed a radically new, positivist and operationalist view that shattered the classical ideal of clarity grounded in Newtonian and Laplacian determinism in space and time. For Bohr and Heisenberg, classical causality had to be relinquished in favor of a framework in which complementary pictures of reality---such as position and momentum---require mutually exclusive experimental arrangements, yet jointly provide a complete account of the real \cite{Bohr1928Nature,Bohr1928Como}.

As Bohr wrote in the proceedings of the Volta Conference in Como in September 1927 \cite{Bohr1928Nature}:
\begin{quote}
\textit{The very nature of the quantum theory thus forces us to regard the space-time co-ordination and the claim of causality, the union of which characterizes the classical theories, as complementary but exclusive features of the description.}
\end{quote}

In the same year, in his seminal article introducing the uncertainty principle, \textit{\"Uber den anschaulichen Inhalt der quantentheoretischen Kinematik und Mechanik}, Heisenberg argued that Laplacian causality was definitively untenable \cite{Heisenberg1927}:
\begin{quote}
\textit{But what is wrong in the sharp formulation of the law of causality, ``When we know the present precisely, we can predict the future,'' it is not the conclusion but the assumption that is false. Even in principle we cannot know the present in all detail. For that reason everything observed is a selection from a plenitude of possibilities and a limitation on what is possible in the future.}
\end{quote}

For Heisenberg, the central problem of quantum causality concerns the tension between, on the one hand, the causal and deterministic evolution of the wave function understood as a catalogue of potentialities (\textit{potentia}) and, on the other hand, the measurement process, which necessarily involves a random phenomenon of collapse and actualization of one of the alternatives contained in this catalogue. This is the origin of the measurement problem, in which the very location of this collapse defines a cut---not only epistemic but ontological---between the quantum and classical worlds. Heisenberg's approach, completed by von Neumann \cite{vonNeumann1932}, and interwoven with Bohr's, constitutes a scaffolding that is not always fully coherent, known as the Copenhagen interpretation---a term Heisenberg himself introduced in 1955 in his Gifford Lectures and popularized in 1958 in \textit{Physics and Philosophy} \cite{Heisenberg1955Gifford,Heisenberg1958}.

Bohr and Heisenberg's approach to causality is, however, far from universally accepted. As early as 1927, at the fifth Solvay Conference, Louis de Broglie---who had extended the wave-particle duality of light to matter in 1923 \cite{deBroglie1924Thesis,deBroglie1923}---presented an alternative formulation, which he called the ``pilot-wave theory'' \cite{deBroglie1927,deBroglie1927Solvay,deBroglie1928SolvayReport}. In this picture, particles follow deterministic trajectories in space-time, as in classical physics, guided by a Schr\"odinger wave propagating in configuration space. Laplacian causality is thus preserved, along with the ontological clarity of the classical world of Newton and Einstein. As de Broglie explained:
\begin{quote}
\textit{...it does not seem to us that there is any reason to renounce believing in the determinism of individual physical phenomena, and it is in this that our conception, which are very similar in other respects to those of Mr Born, appear nevertheless to differ from them markedly.} \cite{deBroglie1928SolvayReport}
\end{quote}

After being heavily criticized and falling into obscurity for decades, de Broglie's approach was rediscovered by David Bohm in 1952 as a ``hidden-variable theory,'' also referred to as the ``causal interpretation'' \cite{Bohm1952I,Bohm1952II}. Interestingly, an early manuscript by Bohm from 1951–52 was titled ``A Causal and Continuous Interpretation of Quantum Theory'' \cite{DrezetStock}. It is now understood that this theory---today universally known as Bohmian mechanics or de Broglie--Bohm theory (dBB) \cite{deBroglieBohmBook1,deBroglieBohmBook1,DurrGoldsteinZanghi1992,Bricmont2016,Durr2020}---reproduces the full range of quantum phenomena, both in the non-relativistic domain and in large parts of quantum field theory, although no consensus has yet been reached on its definitive formulation in the latter regime \cite{Struyve2010,ColinStruyve2015}.

Long relegated to metaphysics, the debate on causality acquired experimental relevance with Bell's inequalities, which enabled direct tests of quantum mechanics against local hidden-variable theories, including local and causal extensions of de Broglie's proposal \cite{Bell1964,Clauser1969}. Experiments have since established the non-local, or more accurately non-locally causal, nature of quantum theory \cite{Aspect1982,FreedmanClauser1972,Hensen2015,Giustina2015,Shalm2015}. As an explicitly non-local theory, dBB theory is therefore a fully legitimate alternative---a natural completion of quantum mechanics without the conceptual vagueness of the orthodox view.\\
\begin{figure}[h]
\begin{center}
        \includegraphics[width=10cm]{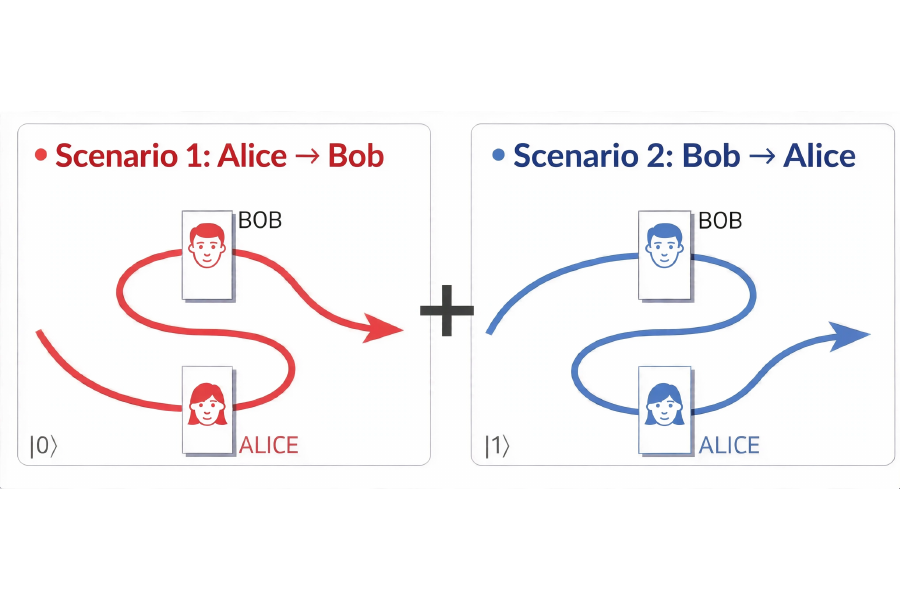}
        \caption{Feynman diagram-style schematic of the quantum switch (QS). Depending on the state of a control qubit, $|0\rangle$ or $|1\rangle$, a target degree of freedom — depicted by the red and blue trajectories — undergoes successive unitary operations in different orders: first Alice then Bob, or first Bob then Alice. Their coherent superposition defines an indefinite causal order (ICO) state}
        \label{fig:0}
  \end{center}      
\end{figure}
\indent  In the last fifteen years, however, the question of causality in quantum mechanics has been radically revived with the introduction of superposed causal states---the so-called indefinite causal order (ICO) \cite{Oreshkov2012,Hardy2007,Hardy2009}. In its original, essentially operational formulation, quantum systems such as qubits or photonic polarization states undergo successive operations in a definite order, say $A$ then $B$ then $C$, or in another order, $C$ then $B$ then $A$ \cite{ChiribellaDArrianoPerinotti2008,Chiribella2013}. Thanks to linearity, quantum mechanics allows superpositions of these alternative causal histories, enabling a host of applications that are in principle impossible with any single, well-defined causal order.

Among them, the quantum switch (QS)---sketched in Fig.~\ref{fig:0}---is the paradigmatic example of ICO. First proposed by Chiribella \textit{et al.} \cite{Chiribella2013}, it is a higher-order quantum operation in which the order of two quantum channels, $\hat{A}$ and $\hat{B}$, is placed in a coherent superposition controlled by a quantum control qubit $|c\rangle$. If $|c\rangle=|0\rangle$, the target system undergoes $\hat{A}$ then $\hat{B}$; if $|c\rangle=|1\rangle$, it undergoes $\hat{B}$ then $\hat{A}$. If the control is prepared in $(|0\rangle+|1\rangle)/\sqrt{2}$, the system experiences $\hat{A}$ then $\hat{B}$ and $\hat{B}$ then $\hat{A}$ in superposition---a causally non-separable process \cite{Oreshkov2012,Branciard2016}. No definite causal history can be assigned. The construction is fundamentally operational: it does not require in principle a classical or quantum space-time, only quantum control of the order of operations performed sequentially \cite{Chiribella2013,ChiribellaWeWeis2013}.

It has been demonstrated photonically with path and polarization as control and target \cite{Procopio2015,Procopio2015NatCommun,Rubino2017SciAdv}, and more recently using time-bin encoding \cite{Rubino2022Quantum,Antesberger2022,Antesberger2024PRL}. Moreover, the landscape of QS implementations and applications has expanded rapidly. Its proven ability to outperform causally ordered circuits in metrology \cite{Zhao2020PRL,Chiribella2012Metrology}, channel discrimination \cite{Bavaresco2022}, communication complexity \cite{Guerin2016,Wei2019}, quantum communication \cite{Ebler2018,Goswami2018,Goswami2020} and thermodynamics \cite{Felce2020,Chiribella2021Thermo} has become a research field in its own right, now covered by several dedicated reviews \cite{Brukner2014,GoswamiReview,ChiribellaReview2021,Rozema2023,Costa2026}.

Reflections on ICO and its paradigmatic example, the QS, are not limited to information processing. Questions of superposed causal structures have, from the outset, lain at the heart of attempts to develop quantum gravity by extending the causal spatio-temporal framework of general relativity into the quantum realm \cite{Hardy2007,Hardy2009,Zych2019,ORiordanHardy2022,Ormrod2023}.

The significance of studies on ICO in recent years---all conducted within a primarily operational framework---nevertheless overlooks the fundamental, ontological questions at the heart of quantum mechanics. Within a Bohmian framework, where the notion of a well-defined trajectory---and thus of history---is central \cite{deBroglieBohmBook1,deBroglieBohmBook2,DurrGoldsteinZanghi1992}, the concept of ICO suggests paradoxes or contradictions that deserve closer examination.

In this article, we explore the connection between ICOs, and the QS in particular, and the causal de Broglie--Bohm theory. We demonstrate how the issue of Bohmian trajectories---which are inherently well-defined and causal---can be reinterpreted in ICO-type scenarios, using the QS as a working example. Ultimately, our main result is that, similar to the elusive wave-particle duality \cite{Bohr1928Nature,Heisenberg1955Gifford}, there is fundamentally no ICO in dBB at the trajectory level. Like any interference effect, ICOs are associated with the guiding wave function, but particles always follow a unique and well-defined history \cite{Bohm1952I,Bohm1952II,deBroglieBohmBook1}.\\
\indent The article is structured as follows. In Section 2, we revisit the principle of thecQs in relation to ICO from an elementary standpoint, using the standard Mach-Zehnder interferometer as a guiding analogy. In Section 3, we briefly recall the theoretical framework of the dBB approach, applied to both spinless and spin- 1/2 particles. In Section 4, we address QS and ICO from a Bohmian perspective and, on the basis of an analytical and numerical model, construct dBB trajectories in QS, thereby showing that QS always admits a description in terms of well-defined paths. In Section 5, we examine the recent problem of the device-independent verification of ICO, as addressed by the no-go theorem of Ref. \cite{vanderLugt2023}, and we argue that, within dBB theory, this theorem cannot rule out the existence of causally well-defined trajectories, owing to the nonlocality inherent in the theory. We conclude with some perspectives for future work
\section{Physics of the quantum switch}\label{sec:ICO}
\begin{figure}[h]
\begin{center}
        \includegraphics[width=10cm]{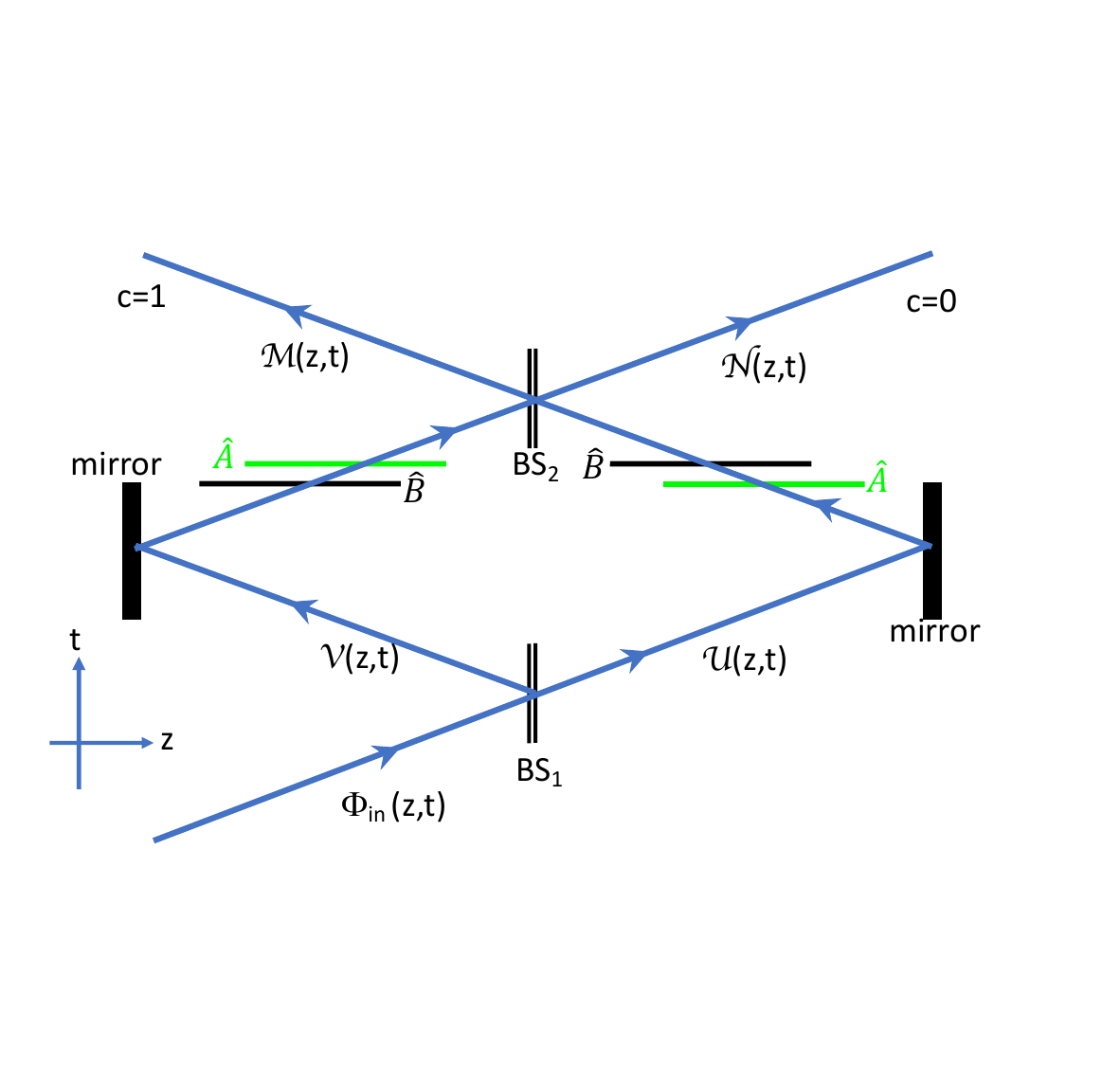} 
        \caption{Schematic diagram of the interferometer acting as an unfolded QS used in our simulations and calculations of dBB trajectories. The system considers, in the $z-t$ plane, two symmetric BSs ($BS_1$ and $BS_2$) and two mirrors operating over a finite time interval so as not to block the propagation of the wave trains, which also have finite spatiotemporal extent. The initial beam splits into two beams, $\mathcal{U}$ and $\mathcal{V}$, and the unitary operators $\hat{A}$ and $\hat{B}$ are placed within the beams to act on the spin degree of freedom (target) associated with the particle. At the end of the experiment, the beams exit the interferometer through outputs $\mathcal{M}$ and $\mathcal{N}$ corresponding to Charlie observable $c=1$ or $c=0$ respectively.}
        \label{fig:1}
        \end{center}
\end{figure}

We begin by reviewing the operating principle of the QS \cite{Chiribella2013,Procopio2015}. To do so, we will limit ourselves to the original implementation used by Procopio et al. \cite{Procopio2015}, which employs an internal degree of freedom (the photon’s polarization in their version) as the “target” and the particle’s path as the “control.” Furthermore, since we are interested in a thought experiment, we will replace the photon and its polarization with a non-relativistic electron and its spin, which has the advantage of having a relatively simple and widely accepted dBB-type description. The implementation of the QS starts with a type of Mach-Zehnder interferometer with unfolded arms, as shown in Fig. \ref{fig:1}. In practice, this diagram is merely a conceptual illustration, and this has led to numerous debates regarding the actual or simulated implementation of the QS \cite{Procopio2015,Goswami2018,PaunkovicVojinovic2020,Abbott2020,Oreshkov2019,Wechs2023,MacLean2017,Zych2019,VilasiniRenner2024,Ormrod2023,Kristjansson2020,Hamette}. However, we will not get bogged down in these debates here and will assume that the QS is perfectly well represented by this “unfolded” version of the QS, in which we have a completely standard Mach-Zehnder interferometer with two well-defined arms, but in which two unitary operations (or gates), $\hat{A}$ and $\hat{B}$, have been placed to act successively on the particle’s spin state. In the two paths, the gates are arranged in the order $\hat{A}$ then $\hat{B}$ or $\hat{B}$ then $\hat{A}$, which generally results in 
different output states, unless the operators commute, i.e., $[\hat{A},\hat{B}]=0$.\\
\indent We write the initial quantum state before the electron entering the QS interferometer as a spinor:
\begin{eqnarray}
  \Psi^{in}(z,t)=  \chi^{in}\Phi^{in}(z,t) 
\end{eqnarray}
where $\Phi^{in}(z,t)\in \mathbb{C}$, `the control' degree of freedom, is a a complex valued and normalized single electron wave  function  belonging to $L^2(\mathbb{R})$ and associated with a localized wave-packet in a 1+1 dimensional space-time $z,t$ (the transverse motion of the wave packet is here neglected). The `target' degree of freedom is the initial pure spinor  
\begin{eqnarray}
\chi^{in}=\left(\begin{array}{c}\chi^{in}_{+z}\\ \chi^{in}_{-z}\end{array}\right)\equiv \chi^{in}_{+z}|+\vec{\mathbf{n}}_z\rangle+\chi^{in}_{-z}|-\vec{\mathbf{n}}_z\rangle
\end{eqnarray} where $\chi^{in}_{\pm z}\in \mathbb{C}$ are the complex probability amplitudes for having respectively a spin up or down along the $z$ direction (we assume the normalization ${\chi^{in}}^\dagger\chi^{in}={\chi^{in}_{+z }}^\ast\chi^{in}_{+z}+{\chi^{in}_{-z }}^\ast\chi^{in}_{-z}=1$).\\
\indent The wave packet associated with the control degree of freedom first encounters a balanced (50/50) beam splitter, denoted $BS_1$ (see Fig.~\ref{fig:1}). The reflected component follows a path in which the particle interacts successively with the unitary gates $\hat{A}$ and then $\hat{B}$, both acting only on the spin (target) degree of freedom. The transmitted component follows the other path, where the same gates are applied in the reverse order, first $\hat{B}$ and then $\hat{A}$.
We denote by $\mathcal{U}(z,t)$ and  $\mathcal{V}(z,t)$ the normalized wave packets associated with the transmitted  and reflected components, respectively, and by $\chi^{B\prec A}= \hat{A}\hat{B}\chi^{in}$ $\chi^{A\prec B}= \hat{B}\hat{A}\chi^{in}$ the corresponding spin states after the action of the unitaries in each branch. Finally, the two wave packets are recombined by a second balanced beam splitter, $BS_2$, yielding the two output normalized wave packets  $\mathcal{N}(z,t)$ and $\mathcal{M}(z,t)$.\\
\indent Formally the different steps of the unitary evolution are summarized by the equation
\begin{eqnarray}
   \Psi^{in}(z,t)\xrightarrow[BS_1]{} \frac{i}{\sqrt{2}}  \chi^{in}\mathcal{V}(z,t)+ \frac{1}{\sqrt{2}}  \chi^{in}\mathcal{U}(z,t)\nonumber\\
   \xrightarrow[mirrors,\hat{A},\hat{B}]{} \frac{-i}{\sqrt{2}}  \chi^{B\prec A}\mathcal{V}(z,t)+ \frac{-1}{\sqrt{2}}  \chi^{A\prec B}\mathcal{U}(z,t)\nonumber\\
   \xrightarrow[BS_2]{}\frac{-1}{2}  (\chi^{A\prec B}-\chi^{B\prec A})\mathcal{M}(z,t)+ \frac{-i}{2}  (\chi^{A\prec B}+\chi^{A\prec B})\mathcal{N}(z,t)\nonumber\\= \frac{-1}{2} [\hat{B},\hat{A}]\chi^{in}\mathcal{M}(z,t)+ \frac{-i}{2}\{\hat{B},\hat{A}\}\chi^{in} \mathcal{N}(z,t)\label{QS1}
\end{eqnarray} where the minus sign is coming from a $\pi-$phase shift due to  wave reflection on the mirrors.
From a computational standpoint, the QS depicted here provides a clear advantage over any causally ordered circuit. As shown by Chiribella et al. \cite{Chiribella2013}, when the unitary gates $\hat{A}$ and $\hat{B}$ either commute or anticommute, the QS can perfectly discriminate between these two cases with a single query to each gate. This is impossible for any conventional sequential circuit with fixed causal order, which requires at least two uses of one of the gates to achieve the same task. The advantage stems directly from the ICO: by placing $\hat{A}$ and $\hat{B}$ in a coherent superposition of the two orderings $\hat{A}\hat{B}$ and $\hat{B}\hat{A}$, the control qubit acquires a relative phase of +1 in the commuting case and -1 in the anticommuting case, making the two alternatives perfectly distinguishable through a simple measurement of the control in the basis associated with the $\mathcal{M}(z,t),\mathcal{N}(z,t)$ spatial modes.\\
\indent Regarding QS, a fundamental question concerns how to quantify its associated ICO— or, more precisely, its causal nonseparability. One approach, developed by analogy with quantum entanglement, introduces a causal witness operator \cite{Branciard2016Witness,Goswami2018,Rubino2017,Araujo2015} that distinguishes causally separable processes (for instance, a configuration involving only one arm of the interferometer) from causally nonseparable ones, such as the QS. This approach ultimately relies on the process matrix formalism, originally developed by Oreshkov, Costa, and Brukner \cite{Oreshkov2012}. Here, we will instead characterize the quality of the QS and its ICO through the Greenberger–Yasin duality inequality\cite{Greenberger1988,Jaeger1995,Englert1996}. We point out that recent works already used this duality in connection with ICO for the QS \cite{Valibouse,Qureshi}. \\ \indent To do this, we first insert a waveplate into one arm of the interferometer to introduce a time delay. This is equivalent to multiplying the probability amplitude in this arm by a phase factor $e^{i\varphi}$. The role of this delay is just to obtain oscillation fringes in the output channels. Moreover, to  further generalize quantum superposition by incorporating environmental effects and decoherence, we assume that the target–control system interacts with external parties, hereafter called Alice and Bob. Alice and Bob receive bipartite inputs x and y, and produce outputs a and b, respectively. The causal order of their interventions differs between the two arms: if, for instance, Alice acts before Bob ($ A \prec B$), she may signal to Bob while he cannot signal to her. Oreshkov, Costa, and Brukner exploited this asymmetry to introduce causal games, such as "Guess Your Neighbor's Input" (GYNI) \cite{Almeida2010GYNI,Branciard2015GYNI,Acin2012GYNI}, which give rise to causal inequalities \cite{Oreshkov2012}. A violation of these inequalities would provide a device-independent certification of ICO. Unfortunately, in Minkowski spacetime the QS does not violate causal inequalities unless one invokes hypothetical quantum superpositions of geometries. Recent work suggests this can be fixed by entangling the QS with a third party and adding extra assumptions — relativistic causality and free interventions — to obtain stronger causal inequalities that the QS can violate. We postpone this to section \ref{VBC} and focus here on the standard model, where Alice and Bob are entangled with the QS, to discuss the Greenberger-Yasin duality inequality. \\
\indent Formally, we here preserve the unitarity of quantum evolution and represent the initial quantum environment $|\mathcal{E}^{\varnothing}\rangle$ as a factorized initial state of the global system $\Psi^{tot}(z,t)=\chi^{in}\Phi^{in}(z,t)|\mathcal{E}^{\varnothing}\rangle$. The quantum evolution of the system through the QS generates two different environmental states $|\mathcal{E}^{A\prec B}\rangle,|\mathcal{E}^{B\prec A}\rangle$ depending on the causal order $A\prec B
$ or $B\prec A$ chosen in the QS. This gives us the following complete quantum state after passing through the first and second beamsplitters:
\begin{eqnarray}
   \Psi^{tot}(z,t)\rightarrow 
    \frac{-i}{\sqrt{2}}  \chi^{B\prec A}\mathcal{V}(z,t)|\mathcal{E}^{B\prec A}\rangle+ \frac{-e^{i\varphi}}{\sqrt{2}}  \chi^{A\prec B}\mathcal{U}(z,t)|\mathcal{E}^{A\prec B}\rangle\nonumber\\
   \rightarrow \frac{-1}{2}  (e^{i\varphi}\chi^{A\prec B}|\mathcal{E}^{A\prec B}\rangle-\chi^{B\prec A}|\mathcal{E}^{B\prec A}\rangle)\mathcal{M}(z,t)+ \frac{-i}{2}  (e^{i\varphi}\chi^{A\prec B}|\mathcal{E}^{A\prec B}\rangle+\chi^{B\prec A}|\mathcal{E}^{B\prec A}\rangle)\mathcal{N}(z,t)\label{QS2}
\end{eqnarray}
which naturally generalizes Eq. \ref{QS1}. The intensities in the output channels read:
\begin{eqnarray}
    I_\mathcal{M}=\frac{1}{4}[\||\mathcal{E}^{A\prec B}\rangle\|^2+\||\mathcal{E}^{B\prec A}\rangle\|^2](1-\mathcal{V}_{A,B}\cos{(\varphi+\varphi_{A,B})})\nonumber\\
    I_\mathcal{N}=\frac{1}{4}[\||\mathcal{E}^{A\prec B}\rangle\|^2+\||\mathcal{E}^{B\prec A}\rangle\|^2](1+\mathcal{V}_{A,B}\cos{(\varphi+\varphi_{A,B})})
\end{eqnarray}
with 
$\langle\mathcal{E}^{B\prec A}|\mathcal{E}^{A\prec B}\rangle
=|\langle\mathcal{E}^{B\prec A}|\mathcal{E}^{A\prec B}\rangle|e^{i\varphi_{A,B}}$ and where the fringe  visibility is defined by
\begin{eqnarray}
    \mathcal{V}_{A,B}=\frac{2|\langle\mathcal{E}^{B\prec A}|\mathcal{E}^{A\prec B}\rangle|\cdot|{\chi^{B\prec A}}^\dagger\chi^{A\prec B}|}{|{\chi^{A\prec B}}^\dagger\chi^{A\prec B}|\cdot\||\mathcal{E}^{A\prec B}\rangle\|^2+|{\chi^{B\prec A}}^\dagger\chi^{B\prec A}|\cdot\||\mathcal{E}^{B\prec A}\rangle\|^2}\nonumber\\=\frac{2|\langle\mathcal{E}^{B\prec A}|\mathcal{E}^{A\prec B}\rangle|}{\||\mathcal{E}^{A\prec B}\rangle\|^2+\||\mathcal{E}^{B\prec A}\rangle\|^2}|{\chi^{B\prec A}}^\dagger\chi^{A\prec B}|
\end{eqnarray} where we used ${\chi^{B\prec A}}^\dagger\chi^{B\prec A}={\chi^{A\prec B}}^\dagger\chi^{A\prec B}={\chi^{in}}^\dagger\chi^{in}$.
Essentially, the fringe visibility $ \mathcal{V}_{A,B}$ (experimentally defined as the ratio  $\frac{I_{max}-I_{min}}{I_{max}-I_{min}}$ in each output channels) measures the degree of quantum coherence between the two histories associated with $A\prec B$ and $B\prec A$. In particular, if we neglect decoherence and can ignore the environment, we obtain the ideal visibility associated with the pure quantum state in Eq. \ref{QS1}:
\begin{eqnarray}
    \mathcal{V}_{A,B}=|{\chi^{B\prec A}}^\dagger\chi^{A\prec B}|=|{\chi^{in}}^\dagger\hat{B}^\dagger\hat{A}^\dagger\hat{B}\hat{A}\chi^{in}|
\end{eqnarray}
In the more general case, the prefactor $\frac{2|\langle\mathcal{E}^{B\prec A}|\mathcal{E}^{A\prec B}\rangle|}{\||\mathcal{E}^{A\prec B}\rangle\|^2+\||\mathcal{E}^{B\prec A}\rangle\|^2}$ represents the environment-related correction to visibility. However, to complete the description of our QS interferometer, we must introduce the concept of distinguishability associated with the particle aspect of our quantum system, according to one of the two alternative histories $A\prec B, B\prec A$. The degree of distinguishability in the QS is trivially defined as:
\begin{eqnarray}
    \mathcal{D}_{A,B}=\frac{|{\chi^{A\prec B}}^\dagger\chi^{A\prec B}|\cdot\||\mathcal{E}^{A\prec B}\rangle\|^2-|{\chi^{B\prec A}}^\dagger\chi^{B\prec A}|\cdot\||\mathcal{E}^{B\prec A}\rangle\|^2}{|{\chi^{A\prec B}}^\dagger\chi^{A\prec B}|\cdot\||\mathcal{E}^{A\prec B}\rangle\|^2+|{\chi^{B\prec A}}^\dagger\chi^{B\prec A}|\cdot\||\mathcal{E}^{B\prec A}\rangle\|^2}\nonumber\\=\frac{|\||\mathcal{E}^{A\prec B}\rangle\|^2-\||\mathcal{E}^{B\prec A}\rangle\|^2|}{\||\mathcal{E}^{A\prec B}\rangle\|^2+\||\mathcal{E}^{B\prec A}\rangle\|^2}.
\end{eqnarray}
In the pure QS of Eq.\ref{QS1} this distinguishability  $ \mathcal{D}_{A,B}$ vanishes meaning that the two arms are playing a symmetrical role.
Crucially, as can be readily shown using the Cauchy–Schwarz inequality, $\mathcal{V}_{A,B}$ and $\mathcal{D}_{A,B}$ satisfy the fundamental Greenberger–Yasin inequality\cite{Greenberger1988,Jaeger1995,Englert1996}:
\begin{eqnarray}
    \mathcal{D}_{A,B}^2+\mathcal{V}_{A,B}^2\leq 1 \label{yasin}
\end{eqnarray}
 In the particular case corresponding to the pure QS of  Eq. \ref{QS1} with $\mathcal{D}_{A,B}=0$
the duality relation reduces to 
\begin{eqnarray}
    \mathcal{V}_{A,B}=|{\chi^{B\prec A}}^\dagger\chi^{A\prec B}|=|{\chi^{in}}^\dagger\hat{B}^\dagger\hat{A}^\dagger\hat{B}\hat{A}\chi^{in}|\leq 1
\end{eqnarray} Furthermore the bound is saturated in the extreme limits corresponding to pure commutation $[\hat{A},\hat{B}]=0$ or anticommutation $\{\hat{A},\hat{B}\}=0$. 
Just as in ``which path" experiments, where precisely determining the path of the particle precludes the simultaneous observation of interference fringes, here a perfect determination of the history followed by the system—for example, $A\prec B$—allows us to obtain the condition $ \mathcal{D}_{A,B}=1$ and  $\mathcal{V}_{A,B}=0$, in accordance with condition \ref{yasin}. We then arrive at a perfectly well-defined causal order. Of course, it is also possible to have a mixture of causal states $A\prec B$ and $B\prec A$ (with $ \mathcal{D}_{A,B}=0$) while still having orthogonal states $\langle\mathcal{E}^{B\prec A}|\mathcal{E}^{A\prec B}\rangle=0$ and thus zero visibility.\\
\indent To illustrate the principle, we examine the GYNI game \cite{Almeida2010GYNI,Branciard2015GYNI,Acin2012GYNI,Oreshkov2012} from a quantum perspective. Alice and Bob, in their respective laboratories, must each guess the outcome of a quantum die roll, with possible outcomes $x=0,1$ for Alice and $y=0,1$ for Bob. Let $a $ denote Alice's guess and $b $ Bob's guess. The game is won if each correctly predicts the other's result. If Alice is causally before Bob, she can in principle send a message to inform him of her result, enforcing the condition $b=x$. Her own guess $a$, however, can only be based on chance — e.g., a quantum coin flip — and will be wrong half the time. The situation is reversed if Bob is causally before Alice. In the first case, where Alice is causally before Bob, suppose she distributes an EPR state between qubits $x$ and $y$. Bob's qubit $y$ and Alice's guess $a$ are then completely independent, and the outcomes are uniformly distributed and random. This allows us to define a global quantum environment state for Alice and Bob as 
\begin{eqnarray}
    |\mathcal{E}_{\pm}^{A\prec B}\rangle=\frac{1}{\sqrt{2}}(|00\rangle_{xb}\pm|11\rangle_{xb})\otimes(\frac{1}{2}|00\rangle_{ya}+\frac{1}{2}|01\rangle_{ya}+\frac{1}{2}|10\rangle_{ya}+\frac{1}{2}|11\rangle_{ya})\
\end{eqnarray}
The sign $\pm$ is arbitrary in our toy model. The model could easily be generalized by including an arbitrary phase $\theta$: $\frac{1}{\sqrt{2}}(|00\rangle_{xb}\pm|11\rangle_{xb}\rightarrow \frac{1}{\sqrt{2}}(|00\rangle_{xb}+ e^{i\theta}|11\rangle_{xb} $ but we will avoid  this issue here.  For the reverse configuration, where Bob is causally before Alice, we can similarly define the global environment state as 
\begin{eqnarray}
   |\mathcal{E}_{\pm}^{B\prec a}\rangle=(\frac{1}{2}|00\rangle_{xb}+\frac{1}{2}|01\rangle_{xb}+\frac{1}{2}|10\rangle_{xb}+\frac{1}{2}|11\rangle_{xb}) \otimes \frac{1}{\sqrt{2}}(|00\rangle_{ya}\pm|11\rangle_{ya})
\end{eqnarray}
Note that depending on the choice of the sign $\pm$, these two states may or may not be orthogonal: 
$\langle \mathcal{E}_{\pm}^{A\prec B}|\mathcal{E}_{\pm}^{B\prec a}\rangle=\frac{1}{8}(1\pm 1)^2$. In the scenario where Alice and Bob interact with the control and target qubits passing through the QS, we therefore obtain a distinguishability  ${\mathcal{D}_{A,B}}_{\pm}=0$, and a fringe  visibility of the form
\begin{eqnarray}
    {\mathcal{V}_{A,B}}_{\pm}=|{\chi^{B\prec A}}^\dagger\chi^{A\prec B}|\frac{1}{8}(1\pm 1)^2
\end{eqnarray}
  which may or may not cancel out (i.e., ${\mathcal{V}_{A,B}}_{-}=0$,  ${\mathcal{V}_{A,B}}_{+}=\frac{1}{2}|{\chi^{B\prec A}}^\dagger\chi^{A\prec B}|$) and subsequently  reveal an ICO state in the QS interferometer.\\
\indent To conclude this section, it is worth noting that the analysis presented here is entirely operational: it is based on an orthodox interpretation of quantum mechanics in which observables such as $\mathcal{V}$ or $\mathcal{D}$ are accessible to observers who are ultimately assumed to be macroscopic and classical.    It is now time to move on to a more ontological and dynamic interpretation of ICOs, starting from a ‘completed’ theory of quantum mechanics, namely the dBB theory.
\section{An introduction to de Broglie Bohm dynamics for spin-$1/2$ particle}\label{sec:dBB}
As briefly reminded in the introduction the de Broglie-Bohm theory (dBB), also known as pilot-wave theory or Bohmian mechanics, is a deterministic completion of quantum mechanics that restores a clear particle ontology \cite{deBroglieBohmBook1,DurrGoldsteinZanghi1992,Bricmont2016,Durr2020}. Its basic hypotheses are: (i) a system of N particles always possesses definite positions $Q_k(t)$ in $3D$ space, (ii) the wave function $ \Psi$ evolves at all times according to the linear Schr\"odinger equation and never collapses, and acts as a physical guiding field, and (iii) the statistical predictions of standard quantum mechanics arise from quantum equilibrium, where the initial particle distribution is $|\Psi|^2$-distributed, a property preserved by the dynamics via `equivariance' \cite{DurrGoldsteinZanghi1992,Durr2020}.\\
\indent In the simplest spinless case for a single particle, the wave function $\psi(\vec{\mathbf{x}},t) \in \mathbb{C}$ obeys the Schr\"odinger equation:
\begin{eqnarray}
   i\hbar \partial_t \psi = -\frac{\hbar^2}{2m}\nabla^2\psi + V\psi 
\end{eqnarray} where $V(\vec{\mathbf{x}},t)$ is an external potential.
Writing the wave function in polar form $\psi = R \exp(iS/\hbar)$ with $R,S$ real, the particle is guided by the velocity field $\mathbf{v}^\psi$ defined by the probability current $\vec{\mathbf{j}}=|\psi|^2\vec{\mathbf{v}}^\psi.$ The guiding equation takes two equivalent forms:
\begin{eqnarray}
    \frac{d\vec{\mathbf{Q}}}{dt} = \vec{\mathbf{v}}^\psi(\mathbf{Q},t) = \frac{\hbar}{m}\text{Im}\frac{\vec{\boldsymbol{\nabla} }\psi}{\psi}\Bigg|_{\vec{\mathbf{x}}=\vec{\mathbf{Q}}(t)} = \frac{1}{m}\vec{\boldsymbol{\nabla} }S(\mathbf{x},t)\Bigg|_{\vec{\mathbf{x}}=\vec{\mathbf{Q}}(t)}
\end{eqnarray}
\begin{figure}[hbtp]\begin{center}
        \includegraphics[width=10 cm]{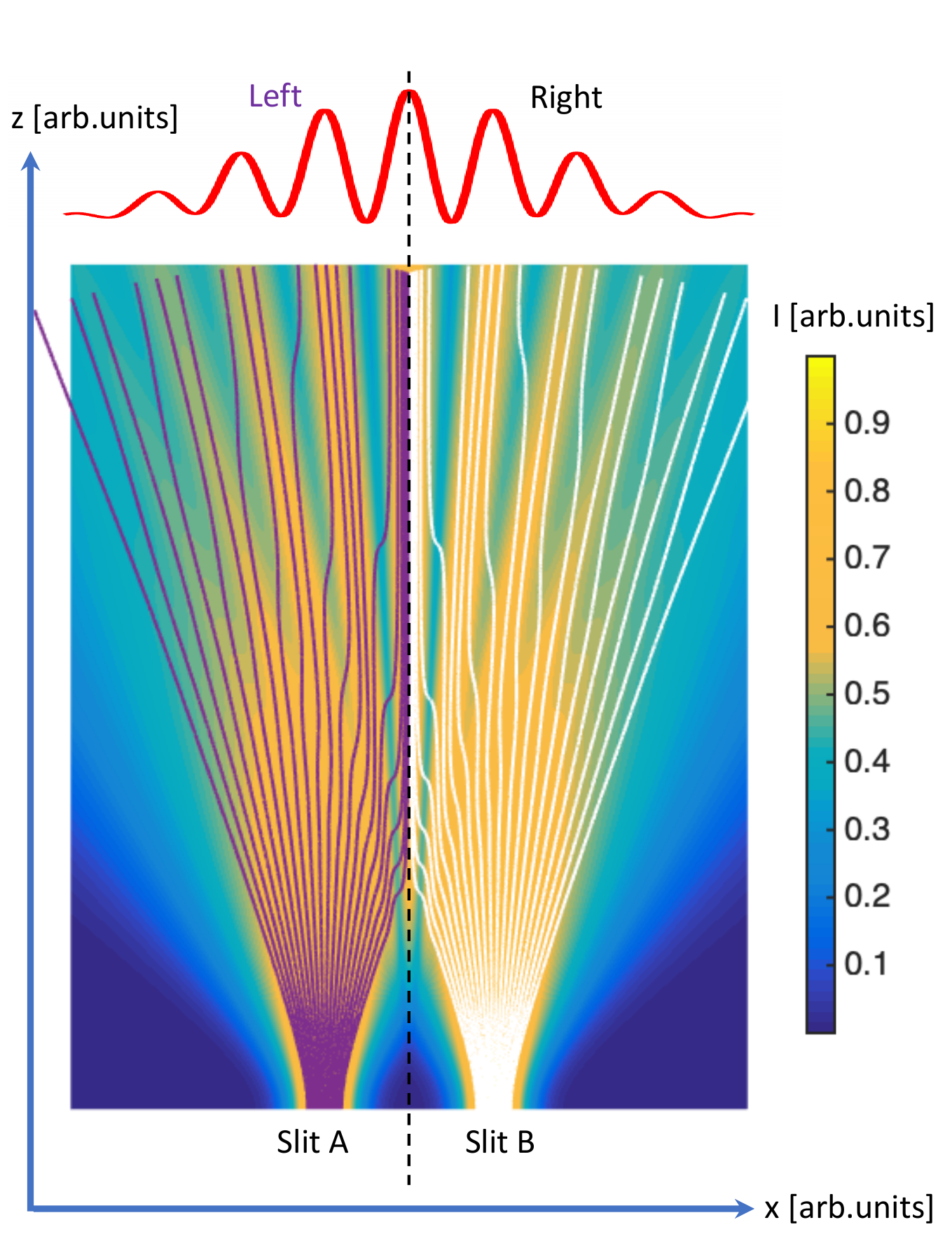}
        \caption{De Broglie-Bohm trajectories computed for an ideal Young double-slit setup in 2D, overlaid on the probability density $\Psi^*\Psi$ in the $x-z$ plane. In dBB theory, trajectories do not intersect, which is reflected here by the existence of two subsets of trajectories (shown in purple and white, respectively) passing through one or the other slit in the left ($x<0$) and right ($x>0$) subspaces, separated by a dashed line. The wave diffracted by the double slit is modeled as $\Psi_+(x,z)+\Psi_-(x,z)$ with  the paraxial  beams $\Psi_\pm(x,z)=\frac{W_0}{W}e^{-\frac{(x\pm d/2)^2}{W^2}}e^{ik(z+\frac{(x\pm d/2)^2}{2R}+\xi)}$, where $W=W_0\sqrt{1+\frac{z^2}{z_0^2}}$, $R=z(1+\frac{z_0^2}{z^2})$, $k\xi=\textrm{atan}(z/z_0)$, $W_0=\sqrt{\frac{\lambda z_0}{\pi}}$ and $k=\frac{2\pi}{\lambda}$. Here we have $W_0=\lambda=1$ and $d=10\lambda$.}
        \label{fig:2}
        \end{center}
\end{figure}
The second form was already introduced by de Broglie in 1927 in analogy with the classical Hamilton-Jacobi theory \cite{deBroglie1927,deBroglie1927Solvay} and by Madelung \cite{madelung} in 1926 in a so-called hydrodynamic picture of the Schr\"odinger equation in analogy with Euler theory for an inviscid and irrotational (potential) fluid. For Madelung, unlike de Broglie, the goal was not to restore the concept of particle trajectory but instead to reformulate Schr\"odinger linear theory as a nonlinear fluid dynamics.
This first-order dynamics (which can be easily generalized to the many body problem) and together with Born's rule $\rho=|\psi|^2$, corresponding to quantum equilibrium, ensures equivariance and reproduces all predictions of standard quantum mechanics at least in the non relativistic regime (there is no consensus in the relativistic regime). \\
\indent One important point which we cannot go into here is that the dBB approach not only constitutes a natural extension of quantum mechanics, utilising its own hydrodynamic formalism (and thus without the addition of any new equations), but it also defines a research programme or framework that makes it possible to extend quantum mechanics beyond its accepted axioms. This is the case, for example, with quantum non-equilibrium approaches which relax the statistical assumptions (Born’s law) without altering the guiding equation. However, one can also utilise Madelung’s hydrodynamic analogy to attempt to modify these dynamics. A review of all these aspects—which we shall not consider further here—is provided in \cite{DrezetEntropy,Valentinibook}.    \\
\indent Moreover, as an illustration of the power of dBB, consider  Young’s double-slit experiment in the Bohmian picture (see a simulation on Fig. \ref{fig:2}). As shown here, we start with two overlapping paraxial Gaussian beams: the dBB trajectories straightforwardly reconstruct the far-field distribution and its interference fringes. This challenges the classical intuition, championed by Bohr and Heisenberg within the Copenhagen interpretation, that such a description is impossible. Importantly — and we will come back to this in the following section — the Bohmian trajectories form a congruence and therefore never intersect. Two distinct families can be clearly identified, each passing through a different slit. This feature is central, as it reveals the underlying history and causal structure of the dBB dynamics.\\
\indent Moreover, for a single non-relativistic spin-$1/2$ particle, this dBB program is implemented through the Pauli equation for the two-component spinor $\Psi(\vec{\mathbf{x}},t) \in \mathbb{C}^2:$
\begin{eqnarray}
i\hbar \partial_t \Psi = \left[ \frac{1}{2m}(-i\hbar\nabla - e\vec{\mathbf{A}})^2 + e\phi - \frac{e\hbar}{2m}\vec{\boldsymbol{\hat{\sigma}}}\cdot\vec{\mathbf{B}} \right]\Psi
\end{eqnarray}
where $\vec{\boldsymbol{\hat{\sigma}}}$ are the Pauli matrices and $\rho = \Psi^\dagger \Psi$ is the density. Here, we introduced the particle charge $e$, the external vector potential $\vec{\mathbf{A}}(\vec{\mathbf{z}} ,t)$, electric potential $\phi(\vec{\mathbf{z}} ,t)$, and magnetic field $\vec{\mathbf{B}}(\vec{\mathbf{z}},t)$ acting on the particle. The particle's trajectory $\vec{\mathbf{Q}}(t)$ is then determined by the associated guiding equation $d\vec{\mathbf{Q}}/dt = \vec{\mathbf{j}}/\rho $ evaluated at $\vec{\mathbf{x}}=\vec{\mathbf{Q}}(t)$, where the Pauli probability current is:
\begin{eqnarray}
\vec{\mathbf{j}} = \frac{\hbar}{2mi}\left(\Psi^\dagger\vec{\boldsymbol{\nabla}}\Psi - (\vec{\boldsymbol{\nabla}}\Psi^\dagger)\Psi\right) - \frac{e}{m}\vec{\mathbf{A}}\rho + \frac{\hbar}{2m}\vec{\boldsymbol{\nabla}}\times(\Psi^\dagger\vec{\boldsymbol{\hat{\sigma}}}\Psi)
\end{eqnarray}
Thus the guiding velocity takes the explicit form \cite{deBroglieBohmBook1,deBroglieBohmBook2,BohmSchillerTiomno1955,BohmSchiller1955,Dewdney}:
\begin{eqnarray}
\frac{d\vec{\mathbf{Q}}}{dt} = \frac{\hbar}{m}\text{Im}\frac{\Psi^\dagger\vec{\boldsymbol{\nabla}}\Psi}{\Psi^\dagger\Psi} - \frac{e}{m}\vec{\mathbf{A}} + \frac{\hbar}{2m}\frac{\vec{\boldsymbol{\nabla}}\times(\Psi^\dagger\vec{\boldsymbol{\hat{\sigma}}}\Psi)}{\Psi^\dagger\Psi} \Bigg|_{\vec{\mathbf{x}}=\vec{\mathbf{Q}}(t)}\label{Pauli}
\end{eqnarray}
In this view, spin is not an intrinsic property carried by the point particle itself, but a property of the guiding spinor field that manifests in the dynamics through both the magnetic term in the Pauli equation and the additional spin-dependent curl term in the guiding equation.\\
\indent Furthermore, it is important to note that the curl term in the probability current is directly obtained from the non-relativistic limit of the Dirac current derived from the relativistic Dirac equation. As pointed out by Holland from relativistic considerations \cite{Holland}, this term is unavoidable in the dBB theory. Still, as already emphasized by Bohm  Schiller and Tiomno, \cite{deBroglieBohmBook1,deBroglieBohmBook2,BohmSchillerTiomno1955,BohmSchiller1955,Dewdney} and Bell \cite{Bell1966}, since the divergence of a curl vanishes, one can define an alternative and simpler Bohmian dynamics by omitting the curl term in Eq.~\ref{Pauli}.
This leads to the minimal dBB guidance formula that we will use in the present work:
\begin{eqnarray}\frac{d\vec{\mathbf{Q}}}{dt} = \frac{\hbar}{m}\text{Im}\frac{\Psi^\dagger\vec{\boldsymbol{\nabla}}\Psi}{\Psi^\dagger\Psi} - \frac{e}{m}\vec{\mathbf{A}}  \Bigg|_{\vec{\mathbf{x}}=\vec{\mathbf{Q}}(t)}\label{PauliBell}\end{eqnarray}
In the present work we will not consider magnetic field and vector potential (i.e., we are not considering Stern-Gerlach apparatus) and we will use the guiding law:
\begin{eqnarray}\frac{d\vec{\mathbf{Q}}}{dt} = \frac{\hbar}{m}\text{Im}\frac{\Psi^\dagger\vec{\boldsymbol{\nabla}}\Psi}{\Psi^\dagger\Psi}  \Bigg|_{\vec{\mathbf{x}}=\vec{\mathbf{Q}}(t)}\label{PauliBellB}\end{eqnarray}
Note that even if the curl term is omitted, an hydrodynamic description analogous to the Madelung fluid equations for spinless particles can still be developed. In this representation, the local spin vector $\vec{\mathbf{S}}=\frac{\hbar}{2}\frac{\Psi^\dagger\vec{\boldsymbol{\hat{\sigma}}}\Psi}{\Psi^\dagger\Psi}$ evolves continuously along the trajectory. In the present work, however, this information is not required and we will restrict ourselves to the guidance formula in Eq.~\ref{PauliBell}.\\
\indent A final remark concerning the rotational term in the probability current: it has recently been suggested that, in certain experiments designed to measure particle arrival times, this term could lead to statistical predictions at odds with standard quantum mechanics. The difficulty stems from the fact that there is no time operator in quantum mechanics — and more generally, no POVM (positive operator-valued measure) associated with the time of detection. Moreover, dBB theory naturally provide a notion of arrival time by integrating the deterministic particle trajectories. For spin-$1/2$ particles in presence of external fields of guiding potentials the dBB trajectories and therefore the arrival time are in general spin-dependent \cite{Das2019}. The Bohmian predictions obtained in this way, however, sometimes violate central theorems concerning POVMs \cite{Goldstein2024}, and no consensus on the interpretation of these results has been reached \cite{Drezet2024,Amblard2026}. Taken at face value, such violations would imply a breakdown of the no-signaling condition and would allow for superluminal information transfer and the possible detection of a preferred foliation of space-time \cite{Drezet2024,Amblard2026}. In the remainder of this paper we will not enter into this debate and will assume that the quantum mechanical predictions hold. This allows us, in the quantum equilibrium regime, to postulate — following Bell — the minimal law Eq.~\ref{PauliBell} as a fundamental principle, in order to provide proofs of principle concerning ICO for QS in what follows.
\section{De Broglie Bohm trajectories in the quantum switch }\label{dBBQS}
In this section, we aim to clarify and decipher how a quantum switch works, demonstrating the emergence of ICO from the perspective of de Broglie-Bohm dynamics. To do this, we will rely on the analysis of the “unfolded” Mach-Zehnder interferometer discussed in Section \ref{sec:ICO} (see Figure \ref{fig:1}).\\
\indent We first consider a single-electron wave packet impinging on a beam splitter (BS).
To simplify the discussion, we consider an incident one-dimensional wave train
along $z$, characterized by the wave function
\begin{eqnarray}
    \Phi^{in}(z,t)\simeq F(z-\beta t)e^{i(k z-\omega t)} \label{num0} 
\end{eqnarray}
where $E:=\hbar\omega=\hbar^2k^2/2m$ is the dispersion relation and 
$\beta=\hbar k/m>0$ is the group velocity.
For mathematical consistency, we further impose $F\simeq \text{const.}=C$ 
within its spatial support where the wave packet does not vanish, and we assume 
$\lambda=2\pi/k \ll \delta z$, where $\lambda$ is the typical wavelength and $\delta z$ 
the typical spatial extension of the packet. Under Born's rule, $|C|^2$ is 
identified with a probability density, so that normalization yields 
$C=1/\sqrt{\delta z}$. The BS is modeled as a rectangular 
potential barrier or well $V(z)=V_0$ with $V_0=\text{const.}$ in the region 
$|z|<\epsilon/2\ll \delta z$ and $V(z)=0$ otherwise.

During the interaction with BS, the total wave function can be 
approximated as
\begin{eqnarray}
\Phi_{BS_1}(z,t)\simeq \Phi^{in}(z,t)+ R \mathcal{V}(z,t) \nonumber \\\textrm{if $z<-\epsilon/2$ }\nonumber \\
    \Phi_{BS_1}(z,t)\simeq F(z-\beta t)[A e^{iq z}+B e^{-iq z}]e^{-i\omega t}\nonumber \\  \textrm{ if $|z|<\epsilon/2$ }\nonumber \\
   \Phi_{BS_1}(z,t)\simeq T\mathcal{U}(z,t)\nonumber \\ \textrm{ if $z>\epsilon/2$ }\label{BS}
\end{eqnarray}
where $\mathcal{U}(z,t)\equiv\psi^{in}(z,t)$ is the transmitted wave train $\mathcal{V}(z,t)\equiv F(x+\beta t)e^{-ik z}e^{-i\omega t}=\psi^{in}(-z,t)$ 
is the reflected wave train, and $R$ (reflection amplitude), $T$ 
(transmission amplitude) and $A,B$ are the Fabry--Perot coefficients 
computed in the limit of an infinitely extended wave packet.

Explicitly, we have
\begin{eqnarray}
T=\frac{4qk}{(q+k)^2}\frac{1}{e^{i(q-k)\epsilon}-\frac{(q-k)^2}{(q+k)^2}e^{-i(k+q)\epsilon}}\nonumber \\
R=iT\frac{k^2-q^2}{2qk}\sin{(q\epsilon)}\nonumber \\
A=R[\frac{q+k}{2q}e^{-i(q-k)\epsilon/2}+\frac{q-k}{2q}e^{-i(q+k)\epsilon/2}]\nonumber \\
B=R[\frac{q-k}{2q}e^{i(q+k)\epsilon/2}+\frac{q+k}{2q}e^{i(q-k)\epsilon/2}]\label{Fabry}
\end{eqnarray}
where $q$ is defined by $E:=\hbar\omega=\hbar^2q^2/2m+V_0$, i.e., 
$q^2-k^2=-2mV_0/\hbar^2$.

For the present work, choosing $\epsilon=\frac{1}{2}\frac{\lambda}{2\pi}$ and 
$q\simeq 2.505\, k_z$ (i.e., $V_0<0$), we obtain 
$T\simeq \frac{1}{\sqrt{2}}e^{i0.267\pi}$ and $R=iT$, which corresponds 
to a balanced $50/50$ BS (with a precision on the BS reflectivity amplitude  of $10^{-5}$).\\
\indent The second stage of our interferometer involves inserting perfectly reflecting mirrors into the paths of the reflected beam $\mathcal{V}$ and the transmitted beam $\mathcal{U}$ after $BS_1.$ Within our idealized model, this corresponds to the interference of the wave $\Phi(z,t)$ defined in Eq. \ref{BS} with two counter-propagating reflected waves. These waves are defined such that the fields u and v vanish at the mirror surfaces, which is equivalent to introducing a $\pi$ phase shift upon reflection.\\
\indent The total wave in the presence of mirrors located at $z=\pm L$, where $2L$ is the spatial width of the interferometer, is then given by:
\begin{eqnarray}
    \Phi(z,t)=\Phi_{BS_1}(z,t)-RF(z+2L-\beta t)e^{ik(z+2L)}e^{-i\omega t}-TF(z-2L+\beta t)e^{-ik(z-2L)}e^{-i\omega t}\label{mirrors}
\end{eqnarray}
Note that in this description, the field for $z<-L$ and $z>L$—located beyond the mirrors—has no physical meaning and corresponds to virtual image fields. Furthermore, since our wave packets are highly localized in both space and time, it is implicit that $BS_1$ acts only during a given period of time centered on the beam, and the same applies to the mirrors, which are assumed to act only when the beams $\mathcal{U}$ and $\mathcal{V}$ arrive at $z=\pm L$.\\
\indent The next step in modeling our QS is to introduce the particle's spin (polarization). $BS_1$ and the mirrors are insensitive to this initial polarization degree $\chi^{in}$, which remains unchanged until the unitary gates $\hat{A}$ and $\hat{B}$ are introduced into the two arms $\mathcal{U}$ and $\mathcal{V}$ of the interferometer after the interaction with the mirrors but before the interaction with $BS_2$. To model all of these processes, we introduce—as for $BS_1$—a field scattered by $BS_2$. This field, which accounts for the delay caused by the fact that the beams in arm $\mathcal{U}$, $\mathcal{V}$ pass through $BS_2$ on average at time $t=2L/\beta$, is given by: 
\begin{eqnarray}
\Phi_{BS_2}(z,t)\simeq \Phi^{in}(z+2L,t)+ R \mathcal{V}(z-2L,t) \nonumber \\\textrm{if $z<-\epsilon/2$ }\nonumber \\
    \Phi_{BS_2}(z,t)\simeq F(z+L-\beta t)[Ae^{iq z}+B e^{-iq z}]e^{i2kL}e^{-i\omega t}\nonumber \\  \textrm{ if $|z|<\epsilon/2$ }\nonumber \\
   \Phi_{BS_2}(z,t)\simeq T\mathcal{U}(z+2L,t)\nonumber \\ \textrm{ if $z>\epsilon/2$ }\label{BS}
\end{eqnarray}
Furthermore, the action of the unitary operators $\hat{A}$ and $\hat{B}$ after interaction with the mirrors, and then taking into account scattering by $BS_2$, is summarized in the following expression, which defines our wave function in the QS, i.e., at time $t >(L+n\delta z)/\beta$ after the interaction with unitaries $\hat{A}$ and $\hat{B}$ (the number $n$ is chosen numerically to avoid overlap with $BS_2$ and the mirrors): 
\begin{eqnarray}
    \Psi(z,t)=\Phi_{BS_1}(z,t)\chi^{in}-R\Phi_{BS_2}(z,t)\hat{A}\hat{B}\chi^{in}-T\Phi_{BS_2}(-z,t)\hat{B}\hat{A}\chi^{in}\label{interQS}
\end{eqnarray}
\begin{figure}[hbtp]\begin{center}
        \includegraphics[width=10cm]{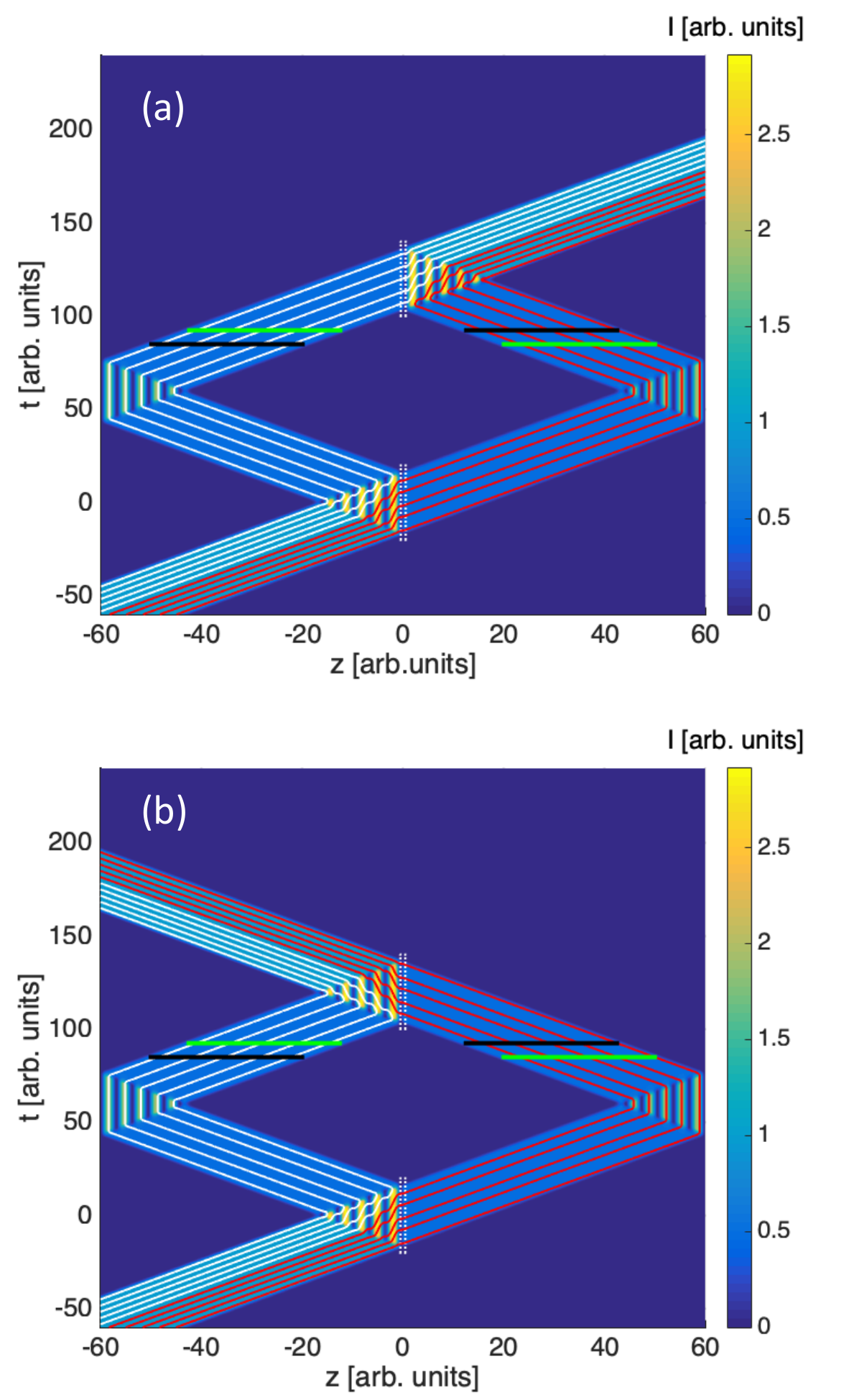} 
        \caption{Bohmian trajectories in the $z-t$ plane for a particle passing through the quantum system schematized in Fig. \ref{fig:1}. The red trajectories follow the path $\mathcal{U}$ associated with the causality $\hat{A}\prec\hat{B}$, while the white trajectories follow the path $\mathcal{V}$ associated with the causality $\hat{B}\prec\hat{A}$.The unitary gates $\hat{A}$ and $\hat{B}$ are $\hat{A}=\hat{B}=i\hat{\sigma}_x$ in case (a) and $\hat{A}=i\hat{\sigma}_x$, $\hat{B}=i\hat{\sigma}_y$ in case (b), respectively. The initial polarization is $\chi^{in}:=|+\vec{\mathbf{n}}_z\rangle$. In each case, an intensity map $\Psi^\dagger\Psi$ in the $z-t$ plane is also shown as a subfigure.}
        
        \label{fig:3}
        \end{center}
\end{figure}
However, using these results, we can model an ideal 1D QS by neglecting beam broadening during propagation through the interferometer. In Figs. \ref{fig:3} and \ref{fig:4}, we present intensity maps in the $z-t$ plane for the QS for different choices of unitary operators $\hat{A}$ and $\hat{B}$ placed in the interferometer arms — corresponding to the causal order $\hat{A}\prec\hat{B}$ for the $\mathcal{U}$ path and $\hat{B}\prec\hat{A}$ for the $\mathcal{V}$ path.  Building on this, and this is the central goal of our modeling, we use the Bohmian mechanics introduced in Sec. \ref{sec:dBB} to numerically compute the dBB trajectories across the entire $z-t $ plane. Based on Eq. \ref{PauliBellB} — here in the absence of a vector potential — these trajectories are calculated for the choices of $\hat{A}$ and $\hat{B}$ shown in Figs. \ref{fig:3} and \ref{fig:4}.\\
\indent As a first example, we consider the cases of perfect commutation $[\hat{A},\hat{B}]=0$ (see Fig. \ref{fig:3}(a)) and perfect anticommutation $\{\hat{A},\hat{B}\}=0$ (see Fig. \ref{fig:3}(b)) of the unitary operators. For a bipartite system, these operators take the general form $\hat{U}=e^{i\frac{\vartheta}{2}\vec{\mathbf{n}}\cdot\boldsymbol{\hat{\sigma}}}=\cos{(\frac{\vartheta}{2})}\mathbb{I}+i\sin{(\frac{\vartheta}{2})}\vec{\mathbf{n}}\cdot\boldsymbol{\hat{\sigma}}$, where $\vartheta\in \mathbb{R}, \vec{\mathbf{n}}\in\mathbb{R}^3$, and $|\vec{\mathbf{n}}|=1$. Specifically, we choose $\hat{A}=\hat{B}=i\hat{\sigma}_x$ for Fig. \ref{fig:3}(a) and $\hat{A}=i\hat{\sigma}_x$, $\hat{B}=i\hat{\sigma}_y$ for Fig. \ref{fig:3}(b). These choices satisfy the required commutation and anticommutation constraints, respectively. The initial polarization is here $\chi^{in}=\left(\begin{array}{c}1\\0\end{array}\right)$ but the result is robust and independent of this choice.\\
\indent Moreover, the focus of this figure is the flow of dBB trajectories in the $z- t$ plane. Since Bohmian dynamics is governed by a first-order differential equation, trajectories cannot cross. As a result, the flow naturally splits into two distinct families, shown in red and white in Fig. \ref{fig:3}. The first family, which follows the path $\mathcal{U}(z,t)$, corresponds to the causal order $\hat{A}\prec\hat{B}$, while the second, which follows $\mathcal{V}(z,t)$, corresponds to $\hat{B}\prec\hat{A}$. This clear separation shows that, at the level of trajectories, there is no causal superposition in dBB theory — even though such a superposition remains present at the level of the guiding wave. In this particular case, an observer, Charlie, familiar with dBB theory who detects a particle in one half of the output beam — beam $\mathcal{M}(z,t)$ in Fig. \ref{fig:3}(a) or beam $\mathcal{N}(z,t)$ in Fig. \ref{fig:3}(b) — corresponding to the red (or white) trajectories, can retrodict its past history and determine whether it followed the causal chain $\hat{A}\prec\hat{B}$ or $\hat{B}\prec\hat{A}$.\\
\begin{figure}[hbtp]\begin{center}
        \includegraphics[width=10cm]{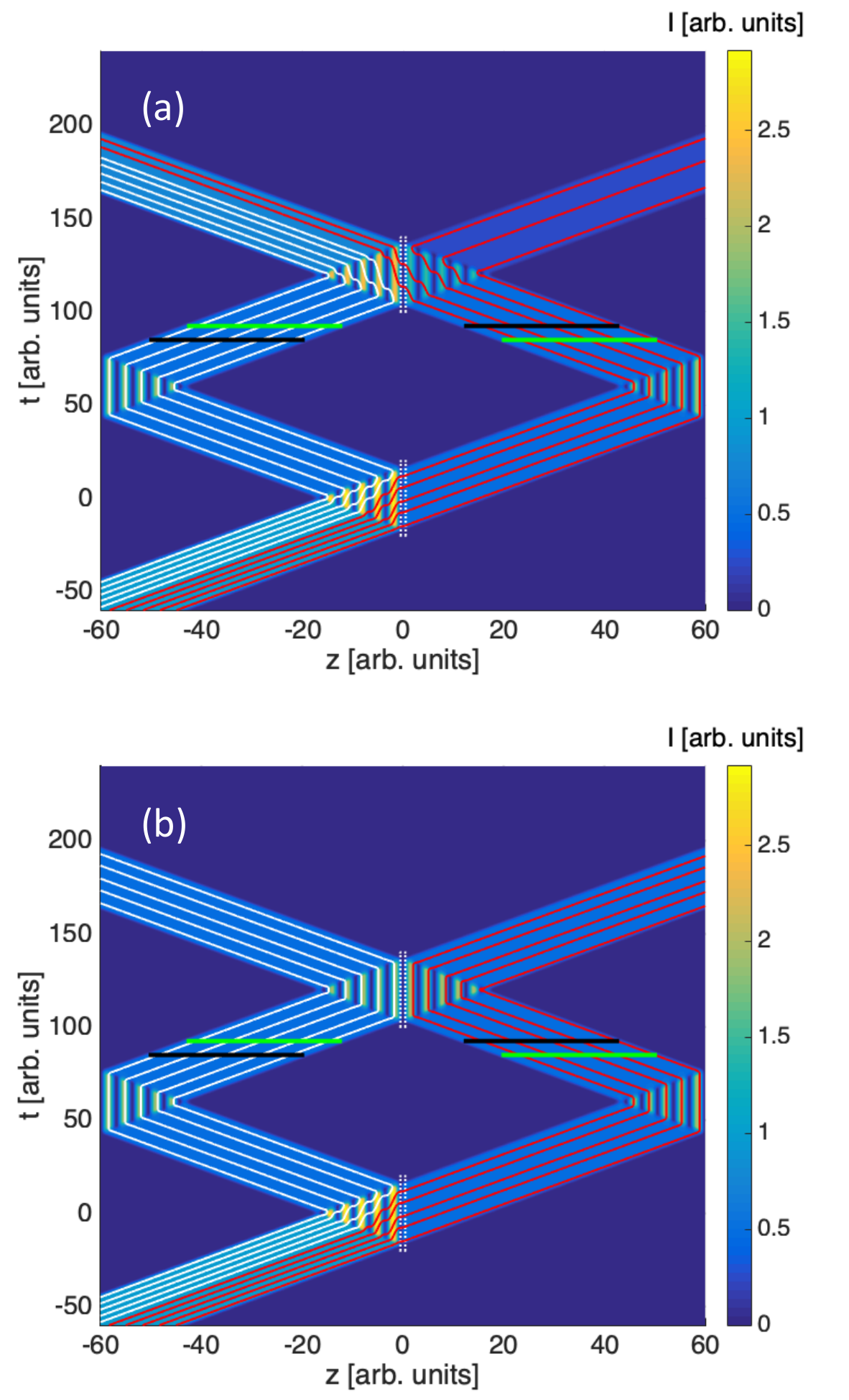}
        \caption{As in Fig. \ref{fig:3}, we show dBB trajectories in the z- t plane for a particle passing through the quantum system sketched in Fig. \ref{fig:1}. The color intensity map and the trajectory color code are the same as in Fig. \ref{fig:3}. Here, the unitary gates are $\hat{A}=i\hat{\sigma}_x$, $\hat{B}=i\frac{\hat{\sigma}_x+\sqrt{3}\hat{\sigma}_y}{2}$ in case (a) and $\hat{A}=i\hat{\sigma}_x$, $\hat{B}=i\frac{\hat{\sigma}_x+\hat{\sigma}_y}{\sqrt{2}}$ in case (b). The initial polarization is again $\chi^{in}:=|+\vec{\mathbf{n}}_z\rangle$.}
        \label{fig:4}
        \end{center}
\end{figure}
\indent The previous analysis is generalizable to any choice of unitary gates $\hat{A},\hat{B}$ in the QS. Consider the intensities $I_\mathcal{M}=\frac{1}{4}\| [\hat{B},\hat{A}]\chi^{in}\|^2$ and $I_\mathcal{N}=\frac{1}{4}\| \{\hat{B},\hat{A}\}\chi^{in}\|^2$ in the output modes $\mathcal{M}$ and $\mathcal{N}$. We can write $I_\mathcal{M}=\frac{1+\varepsilon}{2}$ and $I_\mathcal{N}=\frac{1-\varepsilon}{2}$. Let us consider the case $I_\mathcal{M}-I_\mathcal{N}=\varepsilon\geq 0$. Then, using the continuity of the flux of dBB trajectories through the QS, we can easily deduce that a fraction $\eta=\frac{\varepsilon}{1+\varepsilon}$ of the cross-section of the beam $\mathcal{M}$ comes from the beam $\mathcal{U}$ corresponding to the history $\hat{A}\prec \hat{B}$ with a well-defined causal order. For a rectangular beam $\mathcal{M}$ of width $\sigma$, the separator $x$ — defined relative to the beam center — between the dBB trajectories of type $\hat{B}\prec \hat{A}$ (located on the left side of the output beam $\mathcal{M}$) and those of type $\hat{A}\prec \hat{B}$ (located on the right side) is given by $x=\delta z\left(\frac{1}{2}-\eta\right)$. The remaining trajectories (corresponding to a fraction $1-\eta$) followed the second path $\mathcal{V}$ associated with the causality $\hat{B}\prec \hat{A}$.\\
\indent Of course, a perfectly symmetric situation arises if $\varepsilon\leq 0$. This allows us to make a general inference between the position of the particle detected in the output beam $\mathcal{M}$ or $\mathcal{N}$ and its history—and thus the perfectly well-defined causality (i.e., $\hat{A}\prec\hat{B}$ or $\hat{B}\prec\hat {A}$) associated with each trajectory dBB. In the example in Figure \ref{fig:4}(a), we show a generic case (here obtained for $\hat{A}=i\hat{\sigma}_x$, $\hat{B}=i\frac{\hat{\sigma}_x+\sqrt{3}\hat{\sigma}_y}{2}$). Note that the separation between the trajectories corresponds exactly to our definition of $\eta$. The symmetric case in Figure \ref{fig:4}(b), corresponding to $\hat{A}=i\hat{\sigma}_x$, $\hat{B}=i\frac{\hat{\sigma}_x+\hat{\sigma}_y}{\sqrt{2}}$, is special in that as many trajectories end at $\mathcal{M}$ as at $\mathcal{N}$, and the dBB trajectories completely avoid one another: there is a perfect correlation between, respectively, detecting a particle in $\mathcal{M}$ and following the causality $\hat{B}\prec\hat{A}$, and between detecting a particle in $\mathcal{N}$ and following the causality $\hat{A}\prec\hat{B}$.\\
\indent A central point in our analysis is that, since dBB theory is deterministic, specifying the initial position $z(t_{in})$ of the particle at the initial time (i.e., well before the wave packet interacts with $BS_1$ and the QS) defines a hidden variable $\lambda\equiv z(t_{in})$ in Bell’s sense.  The probabilities for Charlie of observing the particle at output $\mathcal{M}$ (corresponding to Charlie's outcome $c=1$) or $\mathcal{N}$ (corresponding to Charlie's outcome $c=0$) are given by 
\begin{eqnarray}
  \mathcal{P}(c)=\int_\Lambda d\lambda\varrho(\lambda) \mathcal{P}(c|\lambda)
\end{eqnarray}
We have by definition: $\mathcal{P}(1)=I_{\mathcal{M}}$ and $\mathcal{P}(0)=I_{\mathcal{N}}$. Here $\varrho(\lambda)\equiv \Psi(z(t_{in}),t_{in})^\dagger\Psi(z(t_{in}),t_{in})$ defines the hidden variable distribution, and $\mathcal{P}(c|\lambda)\in \{0,1\}$, with $c\in \{0,1\}$, are indicator functions  that depend contextually on the choice of unitary gates $\hat{A},\hat{B}$ located in the arms of the QS  and on the initial polarization state $\chi^{in}$ of the target state (by definition we have $\mathcal{P}(0|\lambda)+\mathcal{P}(1|\lambda)=1$). Indicator functions are given by integrals over distribution functions; for example, $\mathcal{P}(1|\lambda)=\int_ {Z_\mathcal{M}} dz_\mathcal{M}\delta(z_\mathcal{M}-z(t|\lambda,t_{in})$, where $z_c$ is a point in configuration space belonging to the support of the output beam $\mathcal{M}$ with finite support $Z_\mathcal{M}$ in configuration space. $z(t|\lambda,t_{in})$ is a dBB trajectory at time $t$ (where the measure of $\mathcal{M}$ is realized) conditioned on the initial conditions $\lambda$ at time $t_{in}$.\\
\indent Moreover, and as clearly visible from Figs. \ref{fig:3} and \ref{fig:4}, owing to the deterministic character of the dBB dynamics, the space $\Lambda$ of hidden variables $\lambda$ can be divided into two equal halves that correspond exactly to the left and right halves of the initial wave packet of width $\sigma$. This allows us to identify these two halves of $\Lambda$ with the two causal orderings: the right-hand half of the incident beam $\Lambda_{A\prec B}$, associated with the causality $\hat{A}\prec \hat{B}$ and subsequently evolving into the wave $\mathcal{U}$, and the left-hand half $\Lambda_{B\prec A}$ , associated with $\hat{B}\prec \hat{A}$ and subsequently evolving into the wave $\mathcal{V}$.
Taking this separation into account, we write:
\begin{eqnarray}
   \mathcal{P}(c)=\int_{\Lambda_{A\prec B}} d\lambda\varrho(\lambda)\mathcal{P}(c|\lambda) +\int_{\Lambda_{B\prec A}} d\lambda\varrho(\lambda)\mathcal{P}(c|\lambda).
\end{eqnarray} Note that the indicator functions are highly context-dependent and contain information about the operators $\hat{A}, \hat{B}$, but that the division into two causal subspaces arises from the initial distribution.  In general, the probability of observing the Bohmian particle in output channel $c$ or $d$ is therefore always decomposable into a sum of perfectly ordered causal terms.\\
\indent Note that our analysis here is based on a probabilistic decomposition grounded in the configuration space and the concept of hidden variables (or “beables”) $\lambda$. This approach is very simple because it does not require the introduction of measurements by Alice, Bob, or any other agents who are not present in the very simple configuration space associated with our quantum system (here, only the variables associated with the control and the target are present, and in our example, the target is related to spin and is not an essential Bohmian variable, unlike the particle’s position). However, from a more common operationalist perspective in quantum mechanics, the way to view a quantum system or any system exhibiting an ICO as causal is to decompose the probabilities $\mathcal{P}(c,a,b...|z,x,y...)$ associated with measurement outcomes $c,a,b...$ by Charlie, Alice, Bob... into a (convex) sum, $\sum_{h}\mathcal{P}_h(c,a,b|z,x,y)$, of causally well-defined history probabilities. (See, for example, \cite{Araujo2015,Branciard2016Witness}) There are even theorems proving that for quantum systems such as the QS, this is always possible \cite{Branciard2016,Purves2021PRL} and that, in particular, this causal structure implies that quantum mechanics does not allow for the violation of so-called causal inequalities \cite{Oreshkov2012}. It is well known, however, that dBB theory can easily explain and give meaning to the concept of quantum mechanical measurement without all the vagueness and ambiguity of the Copenhagen interpretation. It is therefore worth saying a few words about the connection between our model and the operationalist approach. To this end, the central idea of our Bohmian analysis can easily be generalized to more complex systems comprising multiple degrees of freedom associated with systems coupled to our QS or ICO and involving agents or observers such as Alice, Bob, or Charlie.\\ 
\indent The key point is to realize that any measurement in dBB theory assumes a splitting of the initial wave function—defined in the total configuration space for the entangled system—into non-overlapping subcomponents that are stable in the sense of macroscopic measurement (this is what is called the “effective” collapse of the wave function in dBB theory). This separation makes it possible to give meaning to the notion of outcomes such as $a, b, c...$ conditioned on the parameters $x, y, z...$. The probability of these stable observations is always given by an integral over the hidden variables (the coordinates of the dBB particles in configuration space at the initial instant) and always has a decomposition 
\begin{eqnarray}
    \mathcal{P}(c,a,b...|z,x,y...)=\sum_h\int_{\Lambda_h}
    d\lambda \varrho(\lambda)\mathcal{P}(c,a,b...|\lambda,z,x,y...)
\end{eqnarray}
where $h$ denotes a well-defined causal history corresponding to the dBB trajectories under consideration and having interacted with the measurement systems under consideration. Note that, strictly speaking, the chronology of operations is indicated by pointers associated with clocks that specify which measurements occurred before which others. This will also be achieved by introducing Bohmian variables $\lambda$ and stable observables associated with effective collapses of the wave function. This implies that, strictly speaking, the concept of an ICO does not give rise to any paradoxes in dBB theory. The concept of a causal superposition must therefore be used with caution, just as the concept of wave-particle duality must be. In this regard, it is worth recalling that, according to dBB theory, there is not really a wave-particle duality. At most, one can say, following de Broglie, that there is a harmonious unity between the wave and the particle.  \\ 
\indent Finally a central lesson of our analysis of ICO concerns classical prejudices rooted in naive appeals to `direct measurement', which are all too often invoked in the Copenhagen interpretation to prohibit the very existence of trajectories or determinism, leading to purported paradoxes such as wave-particle duality. In reality, the very idea of direct measurement is illusory. Even Rutherford's experiment does not allow us to deduce a theory; it presupposes one in order to interpret the scattering. As Einstein told Heisenberg in 1926, theory is central to understanding and interpreting quantum mechanics, because it alone defines what can be measured \cite{Heisenberg1969,Heisenberg1971,EinsteinLetter1926}:
\begin{quote}
\textit{it is the theory which alone decides what is measurable.}
\end{quote}
Forgetting this hypothetico-deductive principle is the source of many misinterpretations, as in Young's double-slit experiment. As is well known, dBB trajectories in such a double-slit experiment (see Fig.~\ref{fig:4}) cannot cross --- as is the case for the trajectories in our QS interferometer --- and therefore, by measuring particle positions on a screen where the statistical distribution reconstructs an interferogram, one can, as in our QS setup, retrodict the past history of the dBB particles and obtain which-slit  and interference information without violating the Heisenberg principle \cite{Philippidis1979,Bohm1952I,Bohm1952II,Englert1996,Greenberger1988} .\\
\indent Although this analysis in terms of de Broglie-Bohm trajectories may seem counterintuitive from the perspective of classical physics — which ideally assumes measurements can be performed with arbitrarily high precision and without disturbance — it is certainly less paradoxical than John Wheeler's interpretation of the delayed-choice experiment, according to which one can retroactively reconstruct the past in such a way that \cite{Wheeler1978,Wheeler1984}:
\begin{quote}
\textit{it has no existence except as it is recorded in the present.}
\end{quote}
Wheeler's remark echoes an earlier one by Heisenberg, who stressed as early as 1930 \cite{Heisenberg1930}:
\begin{quote}
\textit{The uncertainty relation does not refer to the past; if the velocity of the electron is at first known and the position then exactly measured, the position for times previous to the measurement may be calculated.}
\end{quote}
\indent In this view, we can know the past but not predict the future. Yet this retrodiction concerns only wave functions, which can be reconstructed retroactively; it does not restore particle trajectories, a concept wholly eliminated by both Heisenberg and Wheeler. The paradoxical consequence of the Copenhagen stance is that the past itself remains undefined until observed. Moreover, the different histories thus reconstructed a posteriori are generally mutually inconsistent and do not coalesce into a single, coherent account of the past. This point is made strikingly clear in Wheeler's delayed-choice examples, which extend the earlier gedanken experiments on retrodiction and indeterminism proposed by von Weizs\"acker and Hermann \cite{vonWeizsacker1931,Hermann1935a,Hermann1935b,Wheeler1978,Wheeler1984}. In contrast, dBB theory restores strict space-time causality and preserves the uniqueness of the system's history, as in classical physics \cite{deBroglie1927,Bohm1952I,Bohm1952II,Bell1987,DurrGoldsteinZanghi1992}.
 All of this is, of course, reinforced in our study of QS and ICO from a Bohmian perspective.   Thus, the history of a particle is strictly preserved in the dBB theory. ICO stems from the concept of interference associated with the pilot wave of the particles, which follow well-defined trajectories such as $\hat{A}\prec \hat{B}$ or $\hat{B}\prec \hat{A}$—never both at the same time.  
\section{Device independent certification of causal order and dBB theory}\label{VBC}
Among the general methods of the operational approach to quantum mechanics, so-called device-independent certification techniques have gained considerable traction. Originally developed for quantum entanglement in the context of quantum cryptography and randomness generation \cite{Acin2007,Pironio2010}, the idea is to treat the systems under study as black boxes defined solely by their inputs and outputs, without any assumption about their internal workings. In the spirit of Bell's analysis of hidden variables \cite{Bell1964,Clauser1969}, one then invokes only general principles---such as locality or free choice---to derive contradictions with a classical worldview. The ultimate goal is to formulate inequalities whose violation certifies that at least one of the axioms required for their derivation must fail. This is precisely what Bell's inequality achieves for nonlocality.

In this context, Oreshkov, Costa and Brukner proposed to extend the device-independent paradigm to causality, by defining causal inequalities whose violation would certify the non-classical nature of indefinite causal order (ICO) \cite{Oreshkov2012,Branciard2016}. The difficulty, however, is that the causal inequalities constructed so far are not violated by quantum theory as physically realized. In particular, no known quantum process has been shown to violate them. This limitation has been formally established: Purves and Short have shown that quantum theory with superposed temporal orders can be simulated by a classical causal model and thus cannot violate any causal inequality \cite{Purves2021PRL}, while Wechs, Dourdent, Abbott and Branciard proved that quantum circuits with quantum control of causal order (QC-QCs)---the class to which the quantum switch belongs---can only generate causal correlations, compatible with a definite causal order \cite{Wechs2021PRXQ,Dourdent2022}.

Due to these difficulties, recent work has introduced a weaker class of causal inequalities based on assumptions strictly stronger than those of Oreshkov, Costa and Brukner. The proposal by van der Lugt, Barrett and Chiribella (hereafter VBC) is to derive Bell-like inequalities that can certify indefinite causal order (ICO) in the quantum switch (QS) provided one assumes both relativistic causality and free interventions \cite{vanderLugt2023,vanderLugt2024,Gogioso}.
In this extended scenario, four parties are considered---Alice 1 and Alice 2 inside the switch, Charlie in their future light-cone, and Bob spacelike separated from the others. Definite causal order, Relativistic causality and Free interventions (DRF) then imply the VBC inequality
\begin{equation}
\mathcal{P}(b=0,a_1=x_2|y=0)+\mathcal{P}(b=1,a_2=x_1|y=0)+\mathcal{P}(b\oplus c=yz|x_1=x_2=1) \leq 7/4,
\label{eq:VBC}
\end{equation}
The first two terms of the VBC inequality form a causal-order game: with a hidden order variable $\lambda$ for the causal order, Bob must guess whether Alice 1 precedes Alice 2 or vice-versa ($b=1,0$), while the Alices attempt to signal to each other ($a_2=x_1$ or $a_1=x_2$). Perfect correlation yields 1 and in the following we will limit the study to  this situation. The third term is a CHSH game between the space-like separated Bob and future Charlie, $b \oplus c = yz$, limited to $3/4$ if Bob's outcome is classically correlated with $\lambda$. The quantum switch violates this trade-off by placing the causal order in superposition: the Alices can signal both ways while the control qubit remains entangled with Bob, allowing $1 + (1/2 + \sqrt{2}/4) \approx 1.85 >1+3/4= 7/4.$ Assuming relativistic causality and free interventions a violation of the VBC inequality therefore rules out any hidden-variable model in which the order between Alice 1 and Alice 2 is fixed by a classical variable. We point out that the DRF axioms are weaker than Bell's local causality usually considered to derive the CHSH  inequality. The VBC inequality inspired by the work of Woodhead on `partially deterministic polytopes' \cite{Woodhead,BarrettKent,Bong,Haddara} presupposes  parameters independence (PI) but not outcomes independences (OI). However due to `partial determinism' (associated with causal order and its hidden variable $\lambda$) the VBC bound $1+3/4$ is indeed justified. Moreover, this VBC  inequality and the observation of the quantum bound $3/2 +\sqrt{2}/4$ have now been tested experimentally  (i.e., in the case where the first two terms sum to $1$) by three independent groups \cite{Richter2025arXiv,Qu2026SciAdv,Guo2026} taking profit on a earlier semi-device-independent demonstration in a photonic switch \cite{Cao2023Optica}, and building on the entanglement-of-order precursor \cite{Rubino2022Quantum}.\\ 
\indent However, such results raise interesting questions when viewed in the context of dBB causal theories, which retain the notion of a trajectory.  We shall therefore now analyse the VBC protocol and its inequality within a Bohmian framework. To do this, we shall draw on an idealised model inspired by our previous calculations with the QS.\\
\indent We start with the spin-singlet Einstein Podolsky Rosen (EPR)  state for a pair of electrons with spatial coordinates $z_A$ and $z_B$ sent respectively to the two Alices and Charlie on the one side and  to Bob on the other side and where $A_1,A_2,C$  and $B$ are space-like separated. The initial state EPR state reads:    
\begin{eqnarray}
    \Psi_{EPR}=\frac{1}{\sqrt{2}}(|+\vec{\mathbf{n}}_z\rangle_A|-\vec{\mathbf{n}}_z\rangle_B-|-\vec{\mathbf{n}}_z\rangle_A|+\vec{\mathbf{n}}_z\rangle_B)\Phi^{in}(z_A)G^{in}(z_B)\label{EPR}
\end{eqnarray}
where the two spatial wave packets $\Phi^{in}(z_A),G^{in}(z_B)$ are located in the two remote labs of the Alices and Charlie on the one side  and Bob on the other side.\\
\indent Here we consider a narrative where Bob measures its particle spin before the particle of the Alices and Charlie (i.e., $B\prec As \prec C$) even reach out the QS interferometer. This presupposes a specific foliation of space-time made of  hyperplanes $\mathcal{F}_{B\prec As \prec C}$ (this defines a Lorentzian rest frame for our description). In the operational picture of quantum  mechanics the role of this foliation is merely accessory and only serves in order to solve the Cauchy problem associated  with the unitary evolution of the full system. The role of  the wave function collapse is nevertheless not very clearly understood in such a relativistic framework, despite many claims, since different foliations imply in general different contradictory narratives. The common wisdom is that this doesn't really matter if one is accepting the Copenhagen interpretation where only probability of outcomes matter in the end. In dBB theory the situation is clearly different. Here the role of the preferred space-time foliation is plainly assumed and the theory is fully consistent. Our choice  $\mathcal{F}_{B\prec As \prec C}$ will therefore implies a specific dBB dynamics and this would very different from a different dBB dynamics using for a example a foliation where Bob measures afters Alices or even Charlie $\mathcal{F}_{As\prec B \prec C}$, $\mathcal{F}_{ As \prec C\prec B}$. We will come back to this issue in the end.\\ 
\indent We first consider the local measurement of the particle spin by Bob when the setting $y=0$ is chosen. In principle, this involves a Stern-Gerlach apparatus, but to remain consistent with our simple dBB dynamics Eq.~\ref{PauliBellB}, which omits magnetic fields and vector potentials, we instead introduce a kind of polarizing beam splitter $PBS_B$. It reflects the spin state $|-\vec{\mathbf{n}}_z\rangle_B$ from the wave packet $G^{in}(z_B)$ into an output mode $G^{b=1}(z_B)$ corresponding to the outcome $b=1$, and similarly transmits the spin state $|+\vec{\mathbf{n}}_z\rangle_B$ into a second output mode $G^{b=0}(z_B)$ — spatially disjoint from $G^{b=1}(z_B)$ — corresponding to the outcome $b=0$. In other words we have the correspondence $|s\mathbf{\hat{z}}\rangle_B\leftrightarrow b=\frac{1-s}{2}$ with $s\in\{-1,+1\}$ and $b\in\{0,1\}$. The quantum state then evolves as:
\begin{eqnarray}
    \Psi_{EPR}\xrightarrow[PBS_B]{}\frac{1}{\sqrt{2}}[|+\vec{\mathbf{n}}_z\rangle_A|-\vec{\mathbf{n}}_z\rangle_BG^{b=1}(z_B)-|-\vec{\mathbf{n}}_z\rangle_A|+\vec{\mathbf{n}}_z\rangle_BG^{b=0}(z_B)]\Phi^{in}(z_A)
    \label{PBSy=0}
\end{eqnarray}
In the next step, the other particle enters the QS. This new QS is built with the same architecture as the one used in the previous Section~\ref{dBBQS}, but with some modified elements. First, $BS_1$ is replaced by a $PBS_A$ acting analogously to $PBS_B$. Second, in the transmitted arm $\mathcal{U}$ (corresponding to the polarization $|+\vec{\mathbf{n}}_z\rangle_A$) we successively place two unitary gates $\hat{A}_1$ and then $\hat{A}_2$ (i.e., associated with the causal order $\hat{A}_1\prec\hat{A}_2$). In the reflected arm $\mathcal{V}$ (corresponding to the polarization $|-\vec{\mathbf{n}}_z\rangle_A$) we first place a `spin-flipper', i.e., a unitary gate $e^{i\frac{\pi}{2}{\sigma_y}_A}$ such that $e^{i\frac{\pi}{2}{\sigma_y}_A}|-\vec{\mathbf{n}}_z\rangle_A=|+\vec{\mathbf{n}}_z\rangle_A$. We then successively insert  the mirrors of QS followed by  two unitary gates $\hat{A}_2$ and then $\hat{A}_1$ (i.e., associated with the causal order $\hat{A}_2\prec\hat{A}_1$). Finally, still in arm $\mathcal{V}$, we insert a phase plate introducing a time delay whose role is to multiply the resulting wave function by a factor $e^{i\pi/2}=i$.
We thus obtain if $y=0$:
\begin{eqnarray}
   \xrightarrow[PBS_A]{}\frac{1}{\sqrt{2}}[|+\vec{\mathbf{n}}_z\rangle_A|-\vec{\mathbf{n}}_z\rangle_BG^{b=1}(z_B)\mathcal{U}(z_A)-|-\vec{\mathbf{n}}_z\rangle_A|+\vec{\mathbf{n}}_z\rangle_BG^{b=0}(z_B)\mathcal{V}(z_A)]\nonumber\\
   \xrightarrow[\text{spin-flipper}]{}\frac{1}{\sqrt{2}}[|+\vec{\mathbf{n}}_z\rangle_A|-\vec{\mathbf{n}}_z\rangle_BG^{b=1}(z_B)\mathcal{U}(z_A)-|+\vec{\mathbf{n}}_z\rangle_A|+\vec{\mathbf{n}}_z\rangle_BG^{b=0}(z_B)\mathcal{V}(z_A)]
   \nonumber\\
   \xrightarrow[mirrors, \hat{A}_2,\hat{A}_1]{}\frac{-1}{\sqrt{2}}[\hat{A}_2\hat{A}_1|+\vec{\mathbf{n}}_z\rangle_A|-\vec{\mathbf{n}}_z\rangle_BG^{b=1}(z_B)\mathcal{U}(z_A)-i\hat{A}_1\hat{A}_2|+\vec{\mathbf{n}}_z\rangle_A|+\vec{\mathbf{n}}_z\rangle_BG^{b=0}(z_B)\mathcal{V}(z_A)]\label{VBC1}
\end{eqnarray}
In the case where $y=1$ we first rewrite  the initial EPR state Eq.\ref{EPR} in the alternative bases $\{|+\vec{\mathbf{n}}_x\rangle_A , |-\vec{\mathbf{n}}_x\rangle_A\}$, $\{|+\vec{\mathbf{n}}_x\rangle_B , |-\vec{\mathbf{n}}_x\rangle_B\}$ associated  with the $\sigma_x$ observable  for the two particles with $|\pm\vec{\mathbf{n}}_x\rangle_i=\frac{1}{\sqrt{2}}[|+\vec{\mathbf{n}}_z\rangle_i\pm |-\vec{\mathbf{n}}_z\rangle_i]$ for $j=A,B$. We have:
    \begin{eqnarray}
    \Psi_{EPR}=\frac{-1}{\sqrt{2}}(|+\vec{\mathbf{n}}_x\rangle_A|-\vec{\mathbf{n}}_x\rangle_B-|-\vec{\mathbf{n}}_x\rangle_A|+\vec{\mathbf{n}}_x\rangle_B)\Phi^{in}(z_A)G^{in}(z_B)\label{EPR2}
\end{eqnarray} 
Measuring the spin of Bob's particle when $y=1$ is very similar to the measurement performed when $y=0$ and involves the use of the same polarizing beam splitter $PBS_B$. However, before the interaction with $BS_B$, we insert a unitary ``spin-rotation'' operator of the form $\hat{U}_{y=1,B}=e^{-i\frac{\pi}{4}\hat{\sigma}_{y_B}}$, which induces the transformations $\hat{U}_{y=1,B} |\pm\vec{\mathbf{n}}_x\rangle_B=\pm|\pm\vec{\mathbf{n}}_z\rangle_B$. The resulting unitary evolution following interaction with this ``spin-rotator'' and then $PBS_B$ yields: 
\begin{eqnarray}
    \Psi_{EPR}\xrightarrow[\hat{U}_{y=1,B},PBS_B]{}\frac{1}{\sqrt{2}}[|+\vec{\mathbf{n}}_x\rangle_A|-\vec{\mathbf{n}}_z\rangle_BG^{b=1}(z_B)+|-\vec{\mathbf{n}}_x\rangle_A|+\vec{\mathbf{n}}_z\rangle_BG^{b=0}(z_B)]\Phi^{in}(z_A)\label{PBSy=1}
\end{eqnarray}
The subsequent evolution is obtained by applying the same unitary operations to the particles sent to the two Alices as in the case $y=0$. The particle interacts with the elements of the QS: first, the $PSB_A$, which splits the beam $\Phi^{in}(z_A)$ into two modes, $\mathcal{U}(z_a)$ and $\mathcal{V}(z_a)$ with different spin states in the computational basis. Then the $\mathcal{V}$ mode undergoes the action of a spin flipper that reverses the spin state from $|-\vec{\mathbf{n}}_z\rangle_A$ to $|+\vec{\mathbf{n}}_z\rangle_A$. This is followed by the introduction of mirrors in the two branches. Subsequently, unitary operations $\hat{A}_1$ and $\hat{A}_2$ are applied to the `target' spin in different sequential orders in the $\mathcal{U}(z_a)$ and $\mathcal{V}(z_a)$ branches. Finally, a phase $\pi/2$ is introduced in beam $\mathcal{V}(z_a)$. This gives us the following resulting state:
\begin{eqnarray}
   \xrightarrow[]{}\frac{1}{\sqrt{2}}\left[\frac{\mathcal{U}(z_A)\hat{A}_2\hat{A}_1+i\mathcal{V}(z_A)\hat{A}_1\hat{A}_2}{\sqrt{2}}|+\vec{\mathbf{n}}_z\rangle_A|-\vec{\mathbf{n}}_z\rangle_BG^{b=1}(z_B)\right.\nonumber\\
   +\left.\frac{\mathcal{U}(z_A)\hat{A}_2\hat{A}_1-i\mathcal{V}(z_A)\hat{A}_1\hat{A}_2}{\sqrt{2}}|+\vec{\mathbf{n}}_z\rangle_A|+\vec{\mathbf{n}}_z\rangle_BG^{b=0}(z_B)\right]\label{VBC2}
\end{eqnarray}

The last step of our idealized experiment consists in recombining the two control beams on a beam splitter $BS$. However, in contrast to the case considered in Section~\ref{dBBQS}, we do not use a symmetric beam splitter, since Charlie, in order to test the CHSH inequality, must consider observables associated with either   $(\hat{\sigma}_z+\hat{\sigma}_x)/\sqrt{2}$ (setting choice $z=0$) or $(\hat{\sigma}_z-\hat{\sigma}_x)/\sqrt{2}$ (setting choice $z=1$) with respect to the initial control qubit basis $\{i\mathcal{V},\mathcal{U}\}$. Physically, this can be implemented by introducing two different $BS_z$ with $z=\{0,1\}$ the setting chosen by Charlie. For $z=0$ we set the reflection and transmission coefficients to $R_{z=0}=i\sin{(\frac{\pi}{8})}e^{i\theta_0}$, $T_{z=0}=\cos{(\frac{\pi}{8})}e^{i\theta_0}$, while for $z=1$ we instead use $R_{z=1}=-i\sin{(\frac{\pi}{8})}e^{i\theta_1}$, $T_{z=1}=\cos{(\frac{\pi}{8})}e^{i\theta_1}$. The phases $\theta_{z}$ are irrelevant here and will be ignored. We stress that, similarly to Section~\ref{dBBQS}, these $BS_z$ can be numerically implemented using Eq.~\ref{Fabry} by choosing $q=1.69504k$ (for $z=0$) and $q=6.54319k$ (for $z=1$), which yield suitable values with a precision on the reflection coefficient of $10^{-6}$.\\
\indent Using $BS_z$, the unitary evolution yields:
\begin{eqnarray}
   \mathcal{U}_{z}= (1-2z)\sin{(\pi/8)}\mathcal{U} +i\cos{(\pi/8)}\mathcal{V}\xrightarrow[BS_z]{} i\mathcal{N}e^{i\theta_z}\nonumber\\
   \mathcal{V}_{z}= -\cos{(\pi/8)}\mathcal{U} +i(1-2z)\sin{(\pi/8)}\mathcal{V}\xrightarrow[BS_z]{} -\mathcal{M}e^{i\theta_z}
\end{eqnarray} 
showing that $\mathcal{U}_{z}$ and $\mathcal{V}_{z}$ indeed act as eigenstates of $BS_z$, i.e., as eigenstates of $(\hat{\sigma}_z+\hat{\sigma}_x)/\sqrt{2}$ for the setting choice $z=0$, or $(\hat{\sigma}_z-\hat{\sigma}_x)/\sqrt{2}$ for $z=1$, in the control qubit basis $\{i\mathcal{V}\equiv\left(\begin{array}{c}1\\0\end{array}\right),\mathcal{U}\equiv\left(\begin{array}{c}0\\1\end{array}\right)\}$.\\
\indent Moreover, after interaction with $BS_z$ in the QS, we obtain if $y=0$ from Eq.~\ref{VBC1} the output quantum state:
\begin{eqnarray}
   \xrightarrow[BS_z]{}\frac{-1}{\sqrt{2}}\hat{A}_2\hat{A}_1|+\vec{\mathbf{n}}_z\rangle_A|-\vec{\mathbf{n}}_z\rangle_BG^{b=1}(z_B)\left(\cos{(\pi/8)}\mathcal{N}(z_A)+i(1-2z)\sin{(\pi/8)}\mathcal{M}(z_A)\right)\nonumber\\+ \frac{1}{\sqrt{2}}\hat{A}_1\hat{A}_2|+\vec{\mathbf{n}}_z\rangle_A|+\vec{\mathbf{n}}_z\rangle_BG^{b=0}(z_B)\left((2z-1)\sin{(\pi/8)}\mathcal{N}(z_A)+i\cos{(\pi/8)}\mathcal{M}(z_A)\right)\label{VBC3}
\end{eqnarray}
If $y=1$ we have similarly from Eq.~\ref{VBC2}:
\begin{eqnarray}
   \xrightarrow[BS_z]{}\frac{1}{2}G^{b=1}(z_B)i\mathcal{N}(z_A)\left((1-2z)\sin{(\pi/8)}\hat{A}_2\hat{A}_1+\cos{(\pi/8)}\hat{A}_1\hat{A}_2\right)|+\vec{\mathbf{n}}_z\rangle_A|-\vec{\mathbf{n}}_z\rangle_B\nonumber\\
   +\frac{1}{2}G^{b=1}(z_B)\mathcal{M}(z_A)\left(\cos{(\pi/8)}\hat{A}_2\hat{A}_1-(1-2z)\sin{(\pi/8)}\hat{A}_1\hat{A}_2\right)|+\vec{\mathbf{n}}_z\rangle_A|-\vec{\mathbf{n}}_z\rangle_B\nonumber\\
   +\frac{1}{2}G^{b=0}(z_B)i\mathcal{N}(z_A)\left((1-2z)\sin{(\pi/8)}\hat{A}_2\hat{A}_1-\cos{(\pi/8)}\hat{A}_1\hat{A}_2\right)|+\vec{\mathbf{n}}_z\rangle_A|+\vec{\mathbf{n}}_z\rangle_B\nonumber\\
   +\frac{1}{2}G^{b=0}(z_B)\mathcal{M}(z_A)\left(\cos{(\pi/8)}\hat{A}_2\hat{A}_1+(1-2z)\sin{(\pi/8)}\hat{A}_1\hat{A}_2\right)|+\vec{\mathbf{n}}_z\rangle_A|+\vec{\mathbf{n}}_z\rangle_B 
   \label{VBC4}
\end{eqnarray}
We now specify the role of the unitaries $\hat{A}_2, \hat{A}_1$ in the VBC setup. In the protocol proposed by VBC \cite{vanderLugt2023}, Alice 1 and Alice 2 in the quantum switch (QS) employ so-called `measure-and-prepare' instruments: for $i=1,2$, Alice $i$ measures the incoming target spin in the computational basis $\{|+\vec{\mathbf{n}}_z\rangle_A, |-\vec{\mathbf{n}}_z\rangle_A\}$ and records the outcome $a_i \in \{0,1\}$. She then re-prepares the target in the state $x_i \in \{0,1\}$, defined in the same computational basis $\{|+\vec{\mathbf{n}}_z\rangle_A, |-\vec{\mathbf{n}}_z\rangle_A\}$.

A rigorous implementation of this protocol requires an evolution operator for the composite system consisting of the target (the spin of particle $A$ traversing the QS) and the memory of each Alice $i$, represented by the state of a qubit $|m_i\rangle_{A_i}$ with $m_i \in \{0,1\}$.

The unitary operators $\hat{A}_i(x_i)$, acting on $\mathcal{H}_{A_i}\otimes\mathcal{H}_A$ and depending on the parameter $x_i$, are directly inspired by \cite{Cao2023Optica,Qu2026SciAdv} and read
\begin{eqnarray}
    \hat{A}_i(x_i)= \hat{P}_i(x_i)\cdot\hat{M}_i(x_i)\label{alicei}
\end{eqnarray}
with
\begin{eqnarray}
    \hat{P}_i(x_i)= |1\rangle_{A_i} {}_{A_i}\langle 1|\hat{F}(x_i)+|0\rangle_{A_i} {}_{A_i}\langle 0|\hat{G}(x_i) \nonumber\\
    \hat{M}_i(x_i)=\hat{\sigma}_{x_{A_i}}|+\vec{\mathbf{n}}_z\rangle_{A} {}_{A}\langle +\vec{\mathbf{n}}_z|+\hat{\mathbb{I}}_{A_i}|-\vec{\mathbf{n}}_z\rangle_{A} {}_{A}\langle -\vec{\mathbf{n}}_z|
\end{eqnarray}
where $\hat{\sigma}_{x_{A_i}}=|1\rangle_{A_i} {}_{A_i}\langle 0|+|0\rangle_{A_i} {}_{A_i}\langle 1|$ and we have introduced the spin rotators $\hat{F}(x_i)=\delta_{x_i,1}\mathbb{I}_A+\delta_{x_i,0}e^{-i\frac{\pi}{2}\hat{\sigma}_{y_A}}$ and $\hat{G}(x_i)=\delta_{x_i,0}\mathbb{I}_A+\delta_{x_i,1}e^{+i\frac{\pi}{2}\hat{\sigma}_{y_A}}$ (we have also added a `A' index to the Pauli matrices $\hat{\sigma}_{x_A}$, $\hat{\sigma}_{y_A}$ acting upon the target degree of freedom A). 
The first factor in chronological order, $\hat{M}_i(x_i)$, corresponds to the measurement stage. It acts as a beam displacer in optics, where a birefringent medium induces a spin-orbit coupling that spatially separates the spin eigenstates $|\pm\vec{\mathbf{n}}_z\rangle_{A}$. The analogy is exact if the qubit of Alice 1, with eigenvectors $\{|0\rangle_{A_1},|1\rangle_{A_1}\}$, is identified with a pointer consisting of two spatially non-overlapping 1D wave packets along the $x$-direction. Similarly, for Alice 2 the eigenvectors $\{|0\rangle_{A_2},|1\rangle_{A_2}\}$ correspond to two non-overlapping 1D wave packets along the $y$-direction. From the perspective of de Broglie-Bohm (dBB) theory, such a physical interpretation is essential, since the structure of configuration space plays a central role in the dynamics.

The second chronological factor, $\hat{P}_i(x_i)$, corresponds to the re-preparation stage, which sets the polarization to the final value $x_i$. Taken together, the resulting beams after the successive action of the operators $\hat{A}_1(x_1)$, $\hat{A}_2(x_2)$ or $\hat{A}_2(x_2)$, $\hat{A}_1(x_1)$ define a point $(a_1,a_2)$ in the $x$-$y$ plane. In our proof of the VBC theorem, we start with the factorizable initial state $\Psi_{EPR}|0\rangle_{A_1}|0\rangle_{A_2}$ and apply the unitary evolutions leading to Eq. \ref{VBC3} for $y=0$ and Eq. \ref{VBC4} for $y=1$, respectively. Importantly, in the present setup the outputs $(a_1,a_2)$ can only take the values $(0,1)$, $(1,0)$ and $(1,1)$. The $(0,0)$ outcome is here forbidden due to the presence of the spin-flipper operator  $e^{i\frac{\pi}{2}\hat{\sigma}_{y_A}}$ reversing the incoming $|-\mathbf{\hat{z}}\rangle_A$ spin state in the $\mathcal{V}$ branch after $PBS_A$. It is important to note that in QS, the causal branch $\hat{A}_1\prec\hat{A}_2$ is associated with the pair $(a_1=1,a_2=x_1)$, and similarly, the causal branch $\hat{A}_2\prec\hat{A}_1$ is associated with the pair $(a_1=x_2,a_2=1)$. The possible values are summarized in Table 1.

\begin{table}
\begin{center}
\begin{tabular}{l||c|c}
  & $x_1=0$ & $x_1=1$\\
\hline
\hline
$x_2=0$  & $\hat{A}_1\prec\hat{A}_2: (a_1=1,a_2=0)$ & $\hat{A}_1\prec\hat{A}_2: (a_1=1,a_2=1)$\\
 & $\hat{A}_2\prec\hat{A}_1: (a_1=0,a_2=1)$ & $\hat{A}_2\prec\hat{A}_1: (a_1=0,a_2=1)$\\
\hline
$x_2=1$  & $\hat{A}_1\prec\hat{A}_2: (a_1=1,a_2=0)$ & $\hat{A}_1\prec\hat{A}_2: (a_1=1,a_2=1)$\\
 & $\hat{A}_2\prec\hat{A}_1: (a_1=1,a_2=1)$ & $\hat{A}_2\prec\hat{A}_1: (a_1=1,a_2=1)$\\
\hline
\end{tabular}
\caption{Table summarizing the possible doublets $(a_1,a_2)$ as a function of $x_1,x_2$.}
\end{center}
\end{table}
The key point of Table 1 is that, for a given choice of $x_1$ and $x_2$, 
the possible outcomes $a_1, a_2$ arising from the two causal orders 
$\hat{A}_1 \prec \hat{A}_2$ and $\hat{A}_2 \prec \hat{A}_1$ generally cannot interfere, as they correspond to distinct output values, i.e., to different points in the configuration space of the spatial pointers of Alice 1 and Alice 2.  We also stress that after the application of $\hat{A}_i(x_i)$ given by Eq.~\ref{alicei} the output target spin-state is necessarily $|(2x_i-1)\vec{\mathbf{n}}_z\rangle_A$ and therefore, just before $BS_z$, the target spin-state is $|(2x_2-1)\vec{\mathbf{n}}_z\rangle_A$ in the $\mathcal{U}$ branch where $\hat{A}_1 \prec \hat{A}_2$, whereas it is given by $|(2x_1-1)\vec{\mathbf{n}}_z\rangle_A$ in the $\mathcal{V}$ branch where $\hat{A}_2 \prec \hat{A}_1$. The only interesting exception is the case $x_1 = x_2 = 1$, which yields $a_1 = a_2 = 1$ in both branches of the QS with the same output polarization $|+\vec{\mathbf{n}}_z\rangle_A$. It therefore corresponds to simple wires, which can interfere at $BS_z$. It is precisely this scenario, $x_1 = x_2 = 1$, that is used in the Bell-CHSH test to demonstrate the violation of the VBC inequality.\\
\indent From Eqs.~\ref{VBC3} and \ref{VBC4}, it is possible to calculate the various probabilities required for the test of the VBC inequality in Eq. \ref{eq:VBC}. For example, due to the perfect correlation obtained in the case $y=0$ between the results $b=0$ and $a_1=x_2$ on the one hand, and $b=1$ and $a_2=x_1$ on the other, we immediately deduce 
\begin{eqnarray}
    \mathcal{P}(b=0,a_1=x_2\mid y=0)=\frac{1}{8}\sum_{x_1,x_2,z\in \{0,1\}}\mathcal{P}(b=0,a_1=x_2\mid x_1, x_2, y=0,z)=\frac{1}{2}
\end{eqnarray} 
Similarly we deduce $\mathcal{P}(b=1,a_2=x_1\mid y=0)=\frac{1}{2}$
which shows that, indeed, the sum of the first two terms of the VBC inequality Eq.~\ref{eq:VBC} is equal to 1. For the term CHSH, it is possible to calculate the joint probability: 
 \begin{eqnarray}
     \mathcal{P}(b\oplus c=yz|x_1=x_2=1)=\frac{1}{4}\sum_{y,z\in \{0,1\}}\mathcal{P}(b\oplus c=yz|x_1=x_2=1,y,z)
 \end{eqnarray}
To do this, we use the joint probabilities $\mathcal{P}(b, c \mid y, z,x_1=x_2=1)$. For example, for $y=z=1$, we obtain 
\begin{align}
\mathcal{P}(b=0, c=1 \mid y=1, z=1,x_1=x_2=1) &= \frac{1}{2} \sin^2\left(\frac{\pi}{8}\right)\nonumber \\
\mathcal{P}(b=0, c=0 \mid y=1, z=1,x_1=x_2=1) &= \frac{1}{2} \cos^2\left(\frac{\pi}{8}\right)\nonumber \\
\mathcal{P}(b=1, c=1 \mid y=1, z=1,x_1=x_2=1) &= \frac{1}{2} \cos^2\left(\frac{\pi}{8}\right)\nonumber \\
\mathcal{P}(b=1, c=0 \mid y=1, z=1,x_1=x_2=1) &= \frac{1}{2} \sin^2\left(\frac{\pi}{8}\right)
\end{align}
which allow us to calculate the average value 
\begin{eqnarray}
    \mathcal{E}_{y=1, z=1}=\mathcal{P}(b=c\mid y=1,z=1, x_1=x_2=1)-\mathcal{P}(b\neq c\mid y=1,z=1, x_1=x_2=1)\nonumber\\=\cos^2\left(\frac{\pi}{8}\right)-\sin^2\left(\frac{\pi}{8}\right)=\frac{1}{\sqrt{2}} 
\end{eqnarray} By doing the same for the other choices of $y$ and $z$, we deduce $\mathcal{E}_{y=0, z=1}=\mathcal{E}_{y=1, z=0}=\mathcal{E}_{y=0, z=0}=-\frac{1}{\sqrt{2}}$ and find the Bell correlation parameter $\mathcal{S}$, which is written as 
\begin{eqnarray}
   \mathcal{S} = |\mathcal{E}_{y=0, z=0} + \mathcal{E}_{y=0, z=1} + \mathcal{E}_{y=1, z=0} - \mathcal{E}_{y=1, z=1}| &= 4 \cdot \frac{1}{\sqrt{2}} = 2\sqrt{2} 
\end{eqnarray}
This allows us, by the construction of the CHSH game, to deduce the probability of winning the game
\begin{eqnarray}
    \mathcal{P}(b \oplus c = y \cdot z\mid x_1=x_2) &= \frac{1}{2} + \frac{\mathcal{S}}{8} = \frac{1}{2} + \frac{\sqrt{2}}{4} \simeq 0.85 
\end{eqnarray} which strongly violates the classical limit of $\mathcal{P}_{\text{cl}} \le \frac{1}{2} + \frac{2}{8} = \frac{3}{4}$ given by the Tsirelson bound. Adding this probability to the other two terms, we indeed obtain the quantum limit $3/2 +\sqrt{2}/4$, which strongly violates the classical limit of $1+3/4=7/4$ given by the VBC inequality Eq.~\ref{eq:VBC}.\\
\indent According to the DRF axioms used to derive the VBC theorem, one might be tempted to conclude that, assuming relativistic causality (parameter independence and the absence of superdeterminism), the idealized experiment described cannot be explained by a hidden-variable model that preserves the notion of causal order in the QS interferometer. \\
\indent However, despite this remarkable result, it is possible to define an explicitly non-local dBB dynamics associated with this quantum system that preserves the notion of causal order at the level of trajectories. To see this, let’s start with the wave function of the entire system, comprising particles A and B with coordinates $z_A(t)$ and $z_B(t)$, as well as the variables of Alice 1 and Alice 2 associated with the discrete coordinates $a_1$ and $a_2$. In reality, these are physically quasi-discrete coordinates, since the particles are assumed to be localized by very sharp wave functions in their respective configuration spaces. The wave function of the coupled system with coordinates $\mathbb{X}:=[z_A, z_B,a_1,a_2]$ in the full configuration space is therefore  formally given by 
\begin{eqnarray}
 \Psi(\mathbb{X},t):= \Psi(z_A, z_B,a_1,a_2, t) &= \sum_{S_A, S_B\in \{-1,+1\}} \gamma_{S_ A, S_B}(a_1,a_2,t)\,
  \chi^{S_A}(z_A, t)\,
  \otimes\phi^{S_B}(z_B, t)   
\end{eqnarray}
where $\chi^{S_A},\phi^{S_B}\in \mathbb{C}^2$ are spinors for the two particles A and B  ($S_A, S_B$ are the spin states along the $z$-direction) and $\gamma_{S_A, S_B}\in \mathbb{C}$ is an amplitude depending on the (quasi-discrete) coordinates of the Alice 1 and Alice 2 pointers. It is easy to see that the wave function of the quantum system previously studied in the context of the VBC theorem is indeed expressed at every time $t$ in this form. It is therefore possible, as in the case of the single particle considered in Section \ref{dBBQS} for the QS, to define a nonlocal  dynamical coupling between the different subsystems. The probability currents for particles A and B are obtained and read as follows: 
\begin{eqnarray}
J_A(z_A, z_B, a_1,a_2, t)=
  \frac{\operatorname{Im}
  \left[ \Psi^\dagger \partial_{z_A} 
  \Psi\right]}{m}\Bigg|_{\mathbb{X}=z_A(t),z_B(t),a_1(t),a_2(t)}\nonumber\\ J_B(z_A, z_B, a_1,a_2, t)=
  \frac{\operatorname{Im}
  \left[ \Psi^\dagger \partial_{z_B} 
  \Psi\right]}{m}\Bigg|_{\mathbb{X}=z_A(t),z_B(t),a_1(t),a_2(t)}
\end{eqnarray}
\begin{figure}[hbtp]\begin{center}
        \includegraphics[width=10cm]{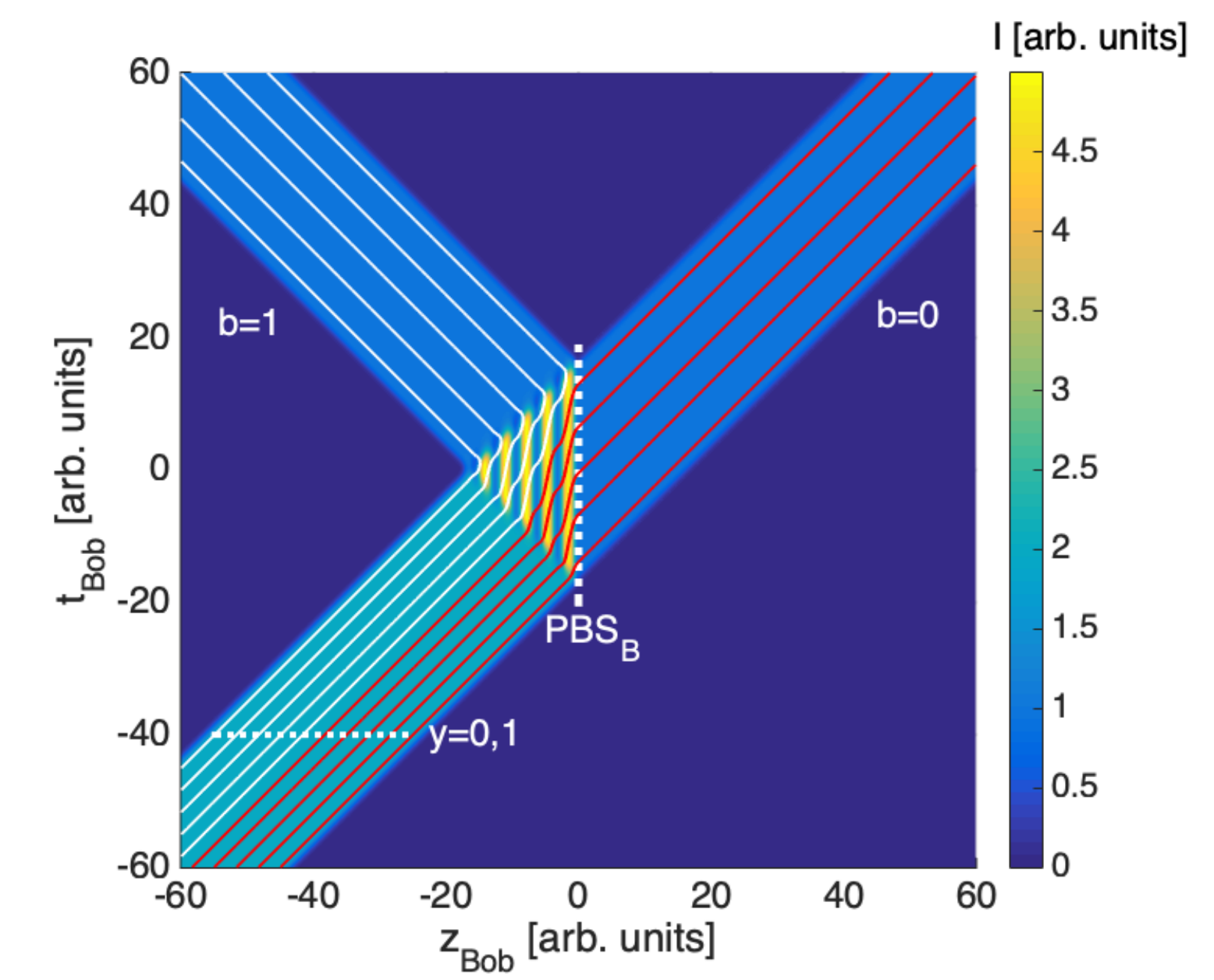}
        \caption{Bohmian trajectories in the $z$-$t$ plane for particle $B$ passing through Bob's $PBS_B$. The transmitted red trajectories end in the beam associated with Bob's outcome $b=0$. Similarly, the reflected white trajectories end in the beam corresponding to Bob's outcome $b=1$. The dynamics is the same irrespective of the setting choice $y\in \{0,1\}$. An intensity map $\Psi^\dagger\Psi$ in the $z$-$t$ plane is also shown as a subfigure. }
        \label{fig:5}
        \end{center}
\end{figure}
The velocity field for the two particles is given by 
\begin{eqnarray}
\frac{d}{dt}z_A(t)= \frac{J_A}{\Psi^\dagger\Psi}\Bigg|_{\mathbb{X}=z_A(t),z_B(t),a_1(t),a_2(t)},&&\frac{d}{dt}z_B(t)= \frac{J_B}{\Psi^\dagger\Psi}\Bigg|_{\mathbb{X}=z_A(t),z_B(t),a_1(t),a_2(t)}
\end{eqnarray} which defines the first order dBB dynamics of the theory.\\
\begin{figure}[hbtp]\begin{center}
        \includegraphics[width=10cm]{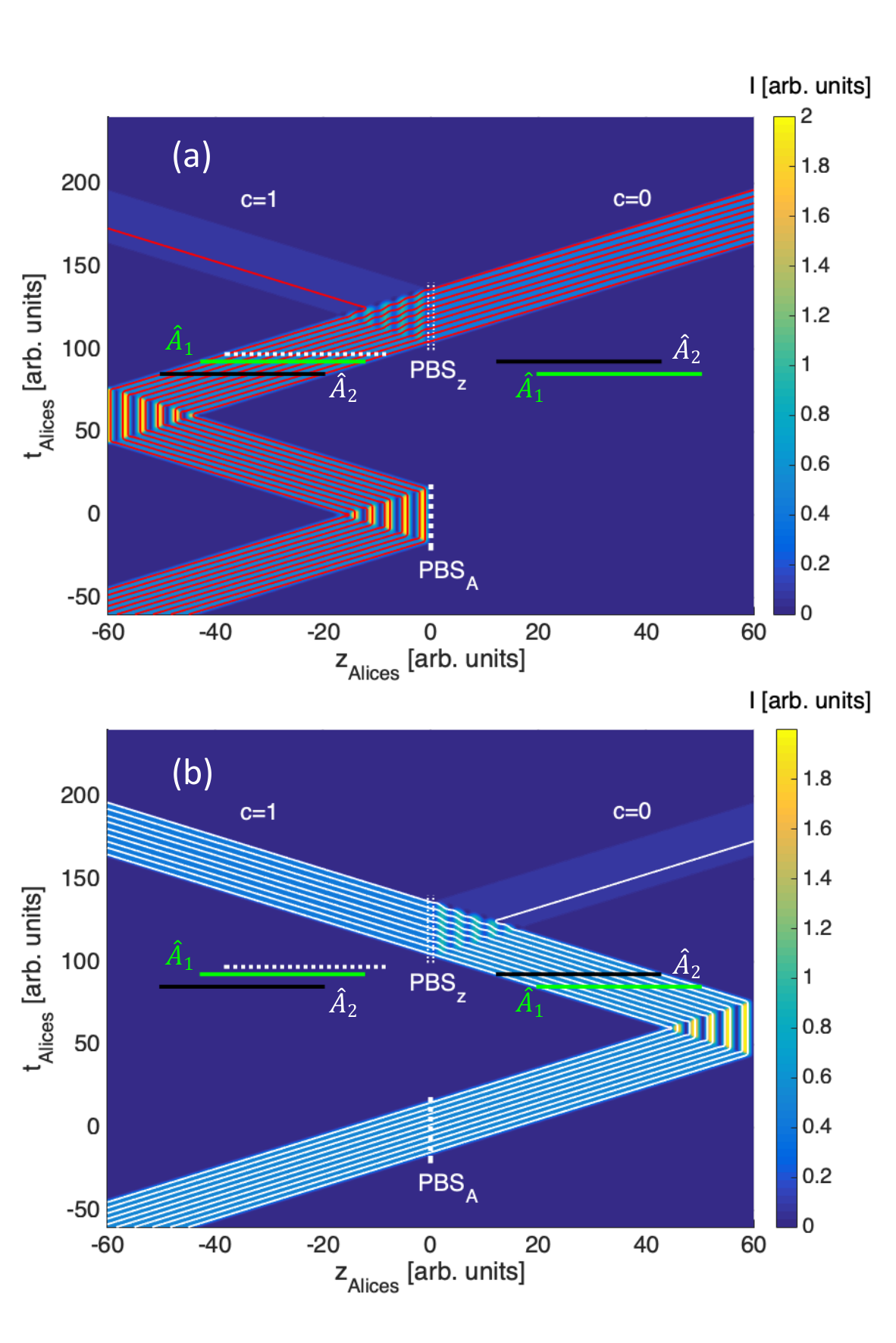}
        \caption{Bohmian trajectories in the $z$-$t$ plane for particle $A$ through the QS in the case $y=0$. The red trajectories in (a) associated  with the causal order $\hat{A}_2\prec\hat{A}_1$, are entangled with $B$ particles following the red paths in Fig.~\ref{fig:5}, corresponding to Bob's outcome $b=0$. Similarly, the white trajectories in (b) associated  with the causal order $\hat{A}_1\prec\hat{A}_2$, are entangled with $B$ particles corresponding to Bob's outcome $b=1$. The dynamics is independent of Charlie's setting choice $z\in \{0,1\}$. In each case, the intensity map $\Psi^\dagger\Psi$ in the $z$-$t$ plane is also shown as a subfigure. }
        \label{fig:6}
        \end{center}
\end{figure}
\begin{figure}[hbtp]\begin{center}
        \includegraphics[width=10cm]{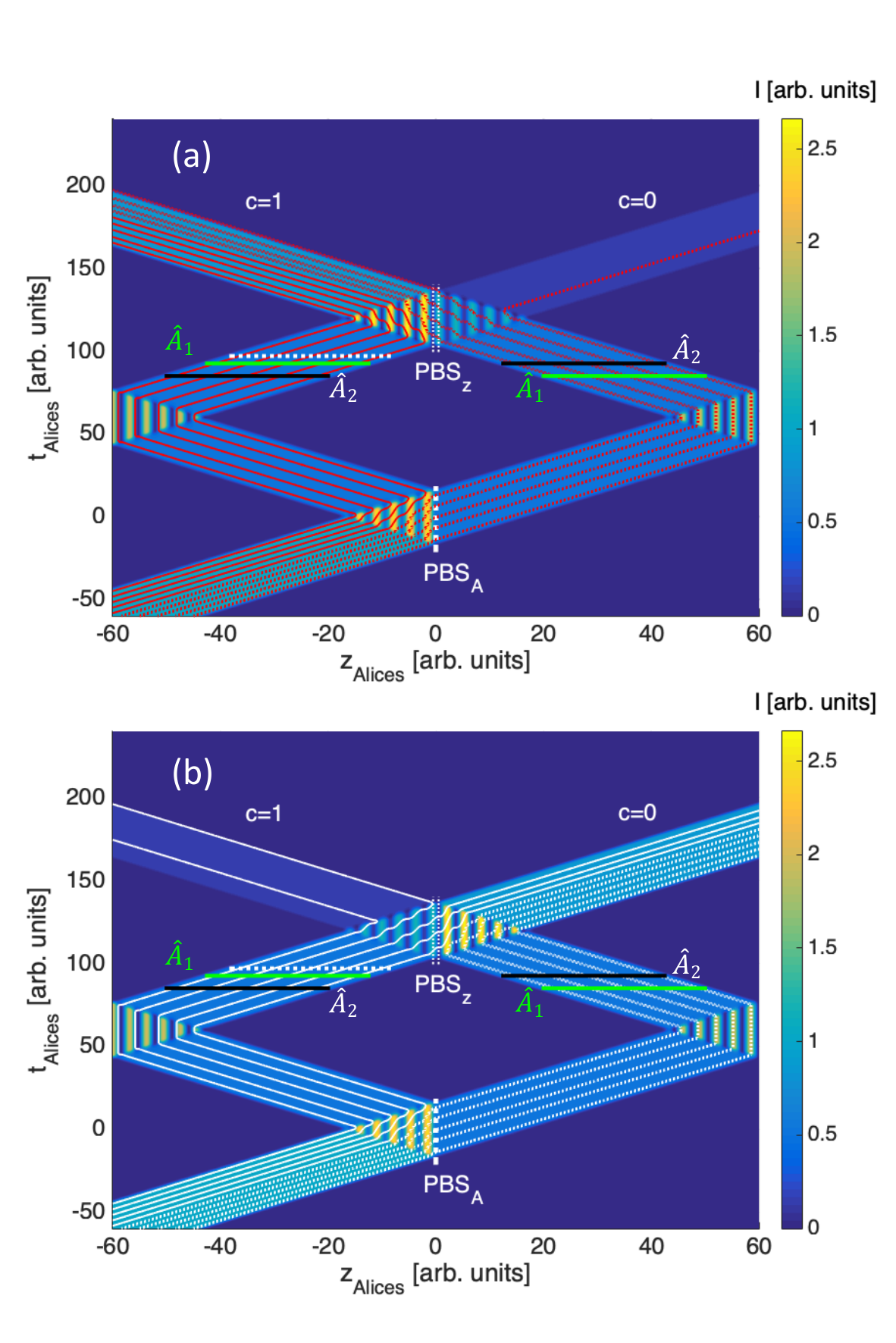}
        \caption{Bohmian trajectories in the $z$-$t$ plane for particle $A$ through the QS in the case $y=1$ and $z=0$ (compare with Fig.~\ref{fig:6}). The red trajectories in (a) are entangled with $B$ particles following the red paths in Fig.~\ref{fig:5}, corresponding to Bob's outcome $b=0$. Similarly, the white trajectories in (b) are entangled with $B$ particles corresponding to Bob's outcome $b=1$. The dotted trajectories are associated with the causal order $\hat{A}_1\prec\hat{A}_2$, whereas the solid ones are associated with the causal order $\hat{A}_2\prec\hat{A}_1$. In each case, the intensity map $\Psi^\dagger\Psi$ in the $z$-$t$ plane is also shown as a subfigure.}
        \label{fig:7}
        \end{center}
\end{figure}
\begin{figure}[hbtp]\begin{center}
        \includegraphics[width=10cm]{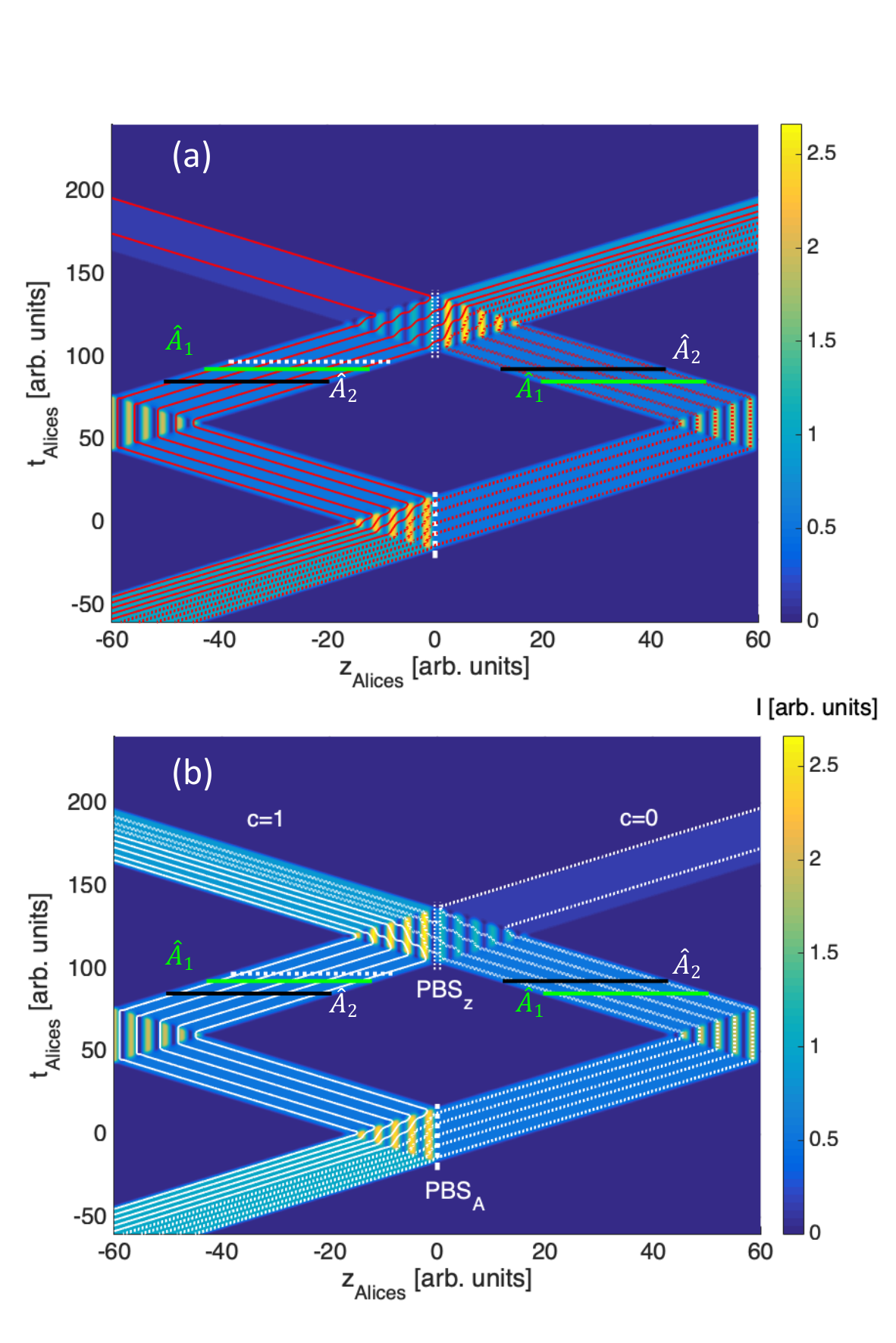}
        \caption{Bohmian trajectories in the $z$-$t$ plane for particle $A$ through the QS in the case $y=1$ and $z=1$. The physical interpretation is the same as for  Fig. \ref{fig:7}.}
        \label{fig:8}
        \end{center}
\end{figure}
\indent Moreover, this dynamics is generally extremely complex and non-linear, owing to the strong non-locality and contextuality inherent to dBB theory. Here, however, the difficulty of the problem is greatly reduced by our choice to consider Bob's and Alice's measurements as time-shifted within our privileged foliation $\mathcal{F}_{B\prec As \prec C}$, in which Bob's measurement on particle $B$ occurs before particle $A$ even enters the QS. In effect, determining Bob's measurement outcome from Eqs.~\ref{PBSy=0} and \ref{PBSy=1} allows us to simplify the description of $B$'s trajectories. Indeed, owing to the initial EPR state, the probability current $J_B$ can be viewed as an equiprobable statistical mixture of the two spin currents associated with $|+\vec{\mathbf{n}}_z\rangle_B$ and $|-\vec{\mathbf{n}}_z\rangle_B$ entering $PBS_B$. This description is, moreover, independent of the value of $y = 0, 1$, since the reduced density matrix of subsystem $B$ is invariant under the change of basis from $|\pm\vec{\mathbf{n}}_z\rangle_B$ to $|\pm\vec{\mathbf{n}}_x\rangle_B$. If $y=1$, the localized spin flipper placed before $PBS_B$, represented by the operator $\hat{U}_{y=1,B}$, therefore has no effect whatsoever on the subsequent dBB dynamics of particle $B$. In this regime, we have rigorously simulated the action of this spin flipper and of $PBS_B$ on the Bohmian dynamics of particle $B$, as shown in Figure~\ref{fig:5}. In this figure, we have plotted the possible Bohmian trajectories of particle B interacting with $PBS_B$, grouped into two subfamilies (shown in white and red, respectively) which, owing to the previously mentioned property of this first-order dynamics, never intersect. They are associated with trajectories terminating in the output beams $b=1$ and $b=0$, respectively.\\
\indent  The dBB trajectories of particles $A$ are entangled with those of $B$.
Since particles $A$ enter the QS well after particles $B$ have left $PBS_B$,
the probability current of $A$ can be written as
\begin{eqnarray}
J_A(z_A, z_B, a_1,a_2, t) &= \sum_{S_A, S_B\in \{-1,+1\}}
  \left|\gamma_{S_A, S_B}(a_1,a_2,t)\right|^2
  {\phi^{S_B}}^\dagger \phi^{S_B}(z_B, t)\,
  \frac{\operatorname{Im}
  \left[ {\chi^{S_A}}^\dagger \partial_{z_A} {\chi^{S_A}}(z_A, t) \right]}{m}
\end{eqnarray}
In this expression, the current of particle $A$ is weighted by the statistical weights of the wave functions of $B$.
Its dependence is different for $y=0$ and for $y=1$.
For $y=0$, the spin state of $A$ is perfectly correlated with Bob's result, as seen in Eq. \ref{PBSy=0}.
As desired, this implies that the red trajectories of $B$ (for $b=0$) shown in Fig. \ref{fig:5} are correlated with those of $A$, also shown in red in Fig.\ref{fig:6}(a), for the motion through the quantum system.
This corresponds to the causal order $\hat{A}_2\prec\hat{A}_1$.
Similarly, $B$'s white trajectories (i.e., for Bob's result $b=1$) are entangled with those shown in white in Fig.\ref{fig:6}(b), which indeed corresponds to the causal order $\hat{A}_1\prec\hat{A}_2$. Note that the dBB trajectories of particle A are the same for $z=0,1$. Note also that the variables $a_1(t)$ and $a_2(t)$ are irrelevant here, as they only introduce additional decoherence to the system's effective wave function, which does not affect the dynamics of particle A in the dBB theory.\\
\indent The case $y=1$ is handled in the same way, but the entanglement between Bob’s particle $B$ and Alice 1’s and Alice 2’s particle $A$ is different and is given by Eq. \ref{PBSy=1}. The correlation between B’s spin $|\pm\vec{\mathbf{n}}_z\rangle_B$ and $A$’s spin $|\mp\vec{\mathbf{n}}_x\rangle_B$ implies that for each value $b=0,1$ of Bob, the effective guiding wave of $A$ passes through both arms of the quantum system at the same time. This implies a form of effective collapse of the wave function of A, and the associated Bohmian dynamics contain, in each case $b=0$ or $b=1$, trajectories that pass through the arm $\mathcal{U}$ associated with the $\hat{A}_1\prec\hat{A}_2$ causality, as well as trajectories that pass through the arm $\mathcal{V}$ associated with the other $\hat{A}_2\prec\hat{A}_1$ causality.\\
\indent In this scenario, unlike in the case $y=0$, we cannot ignore the role of $a_1(t)$ and $a_2(t)$ in the dBB dynamics of particle A. This clearly depends on the choice of the pair $x_1, x_2$. The most interesting case—and the only one we consider here—is $x_1 = x_2 = 1$, which is involved in verifying the CHSH inequality, itself an integral part of the VBC inequality. In this case, we have $a_1=a_2=1$ throughout the entire unitary evolution, which allows us to preserve coherence between the $\mathcal{U}$ and $\mathcal{V}$ branches of the quantum system and thus to perform a test that violates the CHSH inequality. We show in Figs. \ref{fig:7} and \ref{fig:8} the dBB trajectories of particle A in the QS for the cases corresponding to $z=0$ and $z=1$, respectively.   Panels (a) and (b) are, as previously in Fig. \ref{fig:6}, and related to Fig. \ref{fig:5}, which defines the trajectories of particle B through the $PBS_B$ (the correlated trajectories of A and B are shown in red and white, respectively). We can also see two populations of trajectories—solid and dotted lines—that correspond to the causal orders $\hat{A}_2\prec\hat{A}_1$ and $\hat{A}_1\prec\hat{A}_2$ in the QS.\\ 
\begin{figure}[hbtp]\begin{center}
        \includegraphics[width=9 cm]{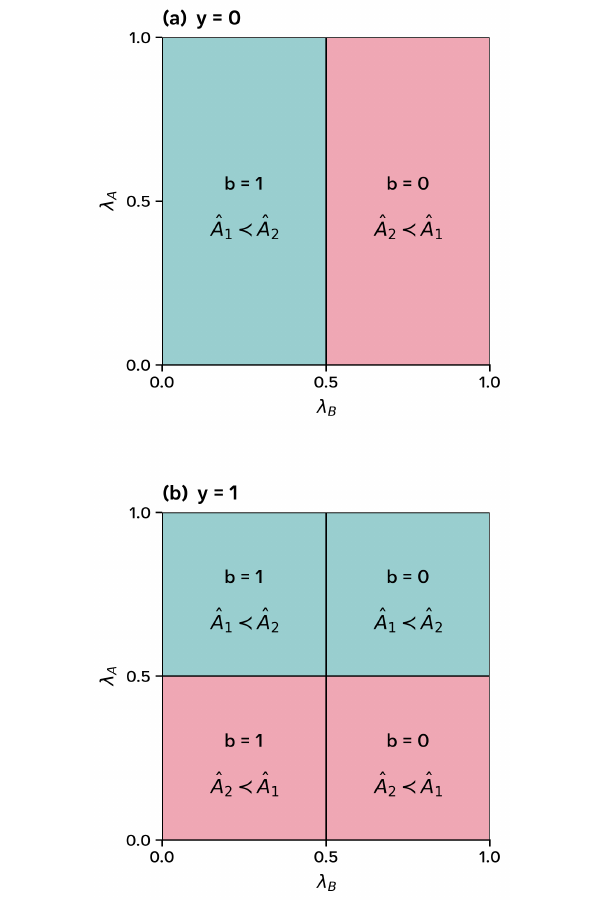}
        \caption{Initial configuration space of B and A particles normalized within a square with $\lambda, \lambda_B \in [0,1]$. Depending on the choice of Bob’s setting—$y=0$ (a) or $y=1$ (b)—the partition of the configuration space into regions corresponding to  $\hat{A}_1\prec \hat{A}_2$ and   $\hat{A}_2\prec \hat{A}_1$ is different. Only case (a) corresponds to a perfect correlation between the causal order and the measurement result of Bob, $b=0,1$. }
        \label{fig:9}
        \end{center}
\end{figure}
\indent Crucially, this shows that the notion of causal order is preserved in the QS even in the presence of quantum entanglement with particle B. What is remarkable here, however, is that in the system under consideration, the VBC inequality is indeed violated to the greatest extent, even though the concepts of dBB trajectories and causal order are perfectly valid and preserved. The crucial question we must therefore ask ourselves is: What does this tell us about the meaning of the VBC inequality and about the ICO device independent certification in QS? What are, in fact, the physical assumptions that are violated, and which, in dBB theory, allow for the preservation of the notion of causal order while, at the same time, are implying a violation of the VBC inequality? \\
\indent If we return to the DRF axioms of the VBC theorem, we naturally think first of relativistic causality, which—due to the nonlocality inherent in dBB—is necessarily affected.  Of course, a violation of the CHSH inequality shows that Alice’s trajectories are nonlocally affected by Bob’s measurements (for the parameter choices $y=0,1$). This is clearly visible in the various figures \ref{fig:5}, \ref{fig:6}, \ref{fig:7}, \ref{fig:8},  as Alice’s trajectories are significantly influenced by Bob’s measurement even when Bob and Alice are spacelike separated. But in fact, these are not the only assumptions that are problematic here. Formally, DRF amounts to assuming the existence of hidden variables $\lambda$ and a well-defined decomposable probability distribution in a causal order superposition : \begin{eqnarray}
   \mathcal{P}(c,b,a_1,a_2\mid z,y, x_1, x_2)=\sum_\lambda \mathcal{P}(\lambda)\mathcal{P}_{A_1\prec A_2}(c,b,a_1,a_2\mid z,y, x_1, x_2,\lambda)\nonumber\\+ \mathcal{P}(\lambda)\mathcal{P}_{A_2\prec A_1}(c,b,a_1,a_2\mid z,y, x_1, x_2,\lambda).
\end{eqnarray}
such that the probabilities for $\hat{A}_1\prec \hat{A}_2$ and $\hat{A}_2\prec \hat{A}_1$ satisfy the condition of parameter independence. This last point is, of course, not satisfied in dBB, but the very idea of a decomposition into non-contextual causal histories can become problematic in dBB due to the nonlocal connection between subsystems A and B. In fact, even if we abandon the parameter independence condition and use the deterministic and nonlocal nature of the dBB theory, we generally do not have the  condition: 
\begin{eqnarray}
   \mathcal{P}_{dBB}(c,b,a_1,a_2\mid z,y, x_1, x_2)=\int_{\Lambda_{A_1\prec A_2}} d\lambda\varrho(\lambda)\mathcal{P}(c,b,a_1,a_2\mid z,y, x_1, x_2,\lambda) \nonumber\\+\int_{\Lambda_{A_2\prec A_1}} d\lambda\varrho(\lambda)\mathcal{P}(c,b,a_1,a_2\mid z,y, x_1, x_2,\lambda),
\end{eqnarray}
where, potentially, the two subspaces $\Lambda_{A_1\prec A_2}$ and $\Lambda_{A_2\prec A_1}$ would be associated with a non-contextual decomposition of the configuration space of the A-B system into two parts corresponding to two well-defined causal orders. The primary explanation stems from the fact that the initial configuration space, in the privileged $\mathcal{F}_{B\prec As \prec C}$ foliation, defines the positions $\lambda_A,\lambda_B$ of the entangled particles before they interact with the measuring devices of the two Alices and Bob. Bob’s measurement, which occurs first, nonlocally affects the subsequent evolution of A in configuration space and the causal order $\hat{A}_1\prec \hat{A}_2$ or $\hat{A}_2\prec \hat{A}_1$ that follows from it. This is clearly visible if we compare Fig. \ref{fig:6} for $y=0$ with the two figures \ref{fig:7} and \ref{fig:8} obtained for $y=1$, in which we see that the distribution of trajectories  that follow the causal orders is affected by Bob’s choice of $y=0$ or $y=1$. To illustrate this, we represent in Figure \ref{fig:9} the initial configuration space as a square with an abscissa normalized to $\lambda_B \in [0,1]$ and an ordinate $\lambda_A \in [0,1]$. This space is contextually and non-locally divided into different causal regions associated with $\hat{A}_1\prec \hat{A}_2$ and  $\hat{A}_2\prec \hat{A}_1$ depending on the choice $y=0$ (Fig. \ref{fig:9} (a)) or $y=1$ (Fig. \ref{fig:9} (b)), in connection with Figures \ref{fig:6}, \ref{fig:7}, and \ref{fig:8}. \\ 
\indent An important point here is that this causal analysis in the dBB approach necessarily presupposes a privileged foliation. Here, the choice $\mathcal{F}_{B\prec As \prec C}$ allows us to demonstrate the contextual nature of causal histories in dBB theory depending on Bob’s choice of the parameter $y=0,1$. In another privileged foliations, the dBB dynamics would be different and would lead to a different partitioning of the configuration space into causal histories. Thus, if in the preferred foliation Alice 1 and Alice 2  take her measurements before Bob, or more simply, if particle A enters the QS before B approaches $PSB_B$ and its spin-flipper, then the dynamics of A would be very different. To compare with Fig. \ref{fig:9}, without going into detail, it is the space $\lambda_A$ that splits into two symmetric parts corresponding to $\hat{A}_2\prec \hat{A}_1$ if $\lambda_A \leq 1/2$ and $\hat{A}_1\prec \hat{A}_2$ if $\lambda_A \geq 1/2$. In this alternative theoretical framework, causality is indeed embedded in the configuration space in a non-contextual manner, and it is the non-local dBB dynamics between A, B, and the pointers that allow us to recover the quantum predictions and thus violate the VBC inequality.\\      
\indent The conclusion of this story is that not only are the locality hypothesis and parameter independence generally irrelevant in quantum mechanics and dBB theory, but also the notion of non-contextual causality is challenged by nonlocality. All of this explains why the dBB theory allows for a violation of the VBC inequality without ever calling into question the idea of a perfectly well-defined causal order. In this context, therefore, the notion of ICO, like that of wave-particle duality, is an illusion stemming from an overly classical, typically Bohrian prejudice. \\
\indent Although such a notion might seem appropriate in a purely instrumentalist approach, it is in itself dangerous and misleading.   In particular, the term ``device-independent,'' so often used in this Bohrian-operationalist context, tends to obscure the fact that, strictly speaking, nothing in physics is device-independent. As Duhem and, above all, Einstein understood very well, a theory is necessary a priori for any description and interpretation of even the slightest empirical data. As Einstein had already told Heisenberg (see above \cite{Heisenberg1969,Heisenberg1971,EinsteinLetter1926}): `it is the theory which alone decides what is measurable'.\\
\indent It is particularly important to remember that as early as 1919 \cite{Einstein1919}, Einstein made the fundamental distinction between “principle-based” theories—such as special relativity or thermodynamics—and “constructive” theories—such as Maxwell’s theory or statistical mechanics—whose more comprehensive structure allows for a more accurate and profound description of reality. In this context, the dBB theory is a constructive theory and stands in opposition to a recent trend known as the “reconstruction program" of the foundations of quantum mechanics, which draws inspiration from principle-based approaches and the device-independent operational approach in quantum information (for a review see \cite{Grinbaum}. Although this operational approach has some value in engineering, it represents a form of marketing for the “second quantum revolution” that cannot be fully justified by its successes. In fact, the dBB approach has the merit of showing that not only is a precise and causal theory of quantum dynamics possible—contrary to many preconceptions that still exist—but that this dBB approach also refutes certain naïve beliefs, accepted too readily, regarding the scope of the device-independent approach in quantum mechanics.
\section{Conclusion and final remarks}
\indent To summarize, in this work we have reexamined the concept of quantum causality — and more specifically, indefinite causal order (ICO) \cite{Oreshkov2012,Hardy2007,Hardy2009} — within the deterministic hidden-variable theory of de Broglie–Bohm (dBB), which preserves the classical notion of particle trajectories, now guided by Schr\"odinger waves. Through the paradigmatic example of the quantum switch (QS), we have shown that causality is not, in fact, undermined by quantum mechanics once it is read through dBB. Contrary to operationalist rhetoric, a Bohmian reading of ICO remains always possible: the phenomenon admits a description in terms of well-defined trajectories unfolding according to well-defined causal orders. Just as wave–particle duality was once invoked by Bohr and Heisenberg as evidence of a supposed “quantum irrationality,” the superposition of causal orders is now invoked to unsettle our classical convictions about objects moving through space and time. In dBB theory, this move fails. Trajectories not only survive, but they reproduce the usual quantum statistics while dissolving the rhetoric of inevitability championed by Bohr and Heisenberg — a rhetoric already fiercely contested by Einstein, de Broglie, Bohm, and Bell.\\
\indent Another point of contention regarding the ICO issue concerns device-independent certification, an approach that has already achieved significant engineering success in quantum key distribution (QKD) and random number generation. Applied to ICO, this approach rests on a few so-called “natural” axioms — Definite Causal Order, Relativistic Causality, and Free Interventions (DRF) \cite{vanderLugt2023,vanderLugt2024,Gogioso} — whose validity, at least in a classical world, seems beyond doubt. Within quantum mechanics, these axioms lead to a VBC inequality, whose violation would, if the DRF axioms were granted, certify an ICO in an entangled quantum system.

The difficulty, however, is that the DRF axioms are neither necessary nor part of dBB theory. The latter reproduces exactly the predictions of quantum mechanics, including the maximal violations of the VBC inequality, without altering in any way the notions of trajectories and of a perfectly well-defined causal order. Since dBB theory can account for all these results, the plausibility of the proposed “device-independent certification” of ICO becomes questionable, especially as its underlying principles stand in tension with quantum nonlocality, already established by Bell’s theorem. The contrast with Bell’s theorem is, in this respect, both instructive and striking. In Bell’s case, the empirical reality of nonlocality is beyond doubt, once a few loopholes are taken into account — most of which can be studied and corrected, with the possible exception of superdeterminism, which would require a conspiracy at the source that is difficult, if not impossible, to rule out completely \cite{Zeilinger3}. It is, for example, remarkable that a superdeterministic interpretation of dBB theory is possible, building on the time-symmetric approach proposed by de Broglie in connection with the ‘double solution’ theory\cite{Drezet2023,Drezet2024}.  Moreover the verification of nonlocality thus makes it possible to test quantum key distribution for cryptography, whose validity and effectiveness are limited, at most, by the precision of quantum mechanics itself — in particular by Born’s rule and quantum equilibrium in dBB theory.\\
\indent The situation is very different for ICO and its associated device-independent certification. Here, one can always appeal to nonlocality and quantum contextuality — both integral to dBB theory — to preserve a perfectly well-defined causal order. In this framework, the concept of ICO once again appears as a mirage, despite all the advantages it promises for engineering and information technology \cite{Brukner2014,GoswamiReview,ChiribellaReview2021,Rozema2023,Costa2026}.\\
\indent One final remark concerns quantum gravity, often associated with ICO in recent speculative research \cite{Hardy2007,Hardy2009,Brukner2014,Zych2019,VilasiniRenner2024,Hamette,Letertre2022,HamettePRL}. It is remarkable that a Bohmian version of quantum gravity, based on the Wheeler--DeWitt equation, has existed in the literature for a very long time \cite{deBroglieBohmBook2,Valentini1996,PintoNetoStruyve2018,Valentini2023}. In this theory, a dBB guidance formula can be defined for the time derivative $\partial_tg_{ij}(x)$ of the induced 3-metric on a spacelike hypersurface \cite{deBroglieBohmBook2}. Although there is no consensus on the validity of this theory, on its mathematical consistency, or even on the interpretation of its time parameter, it preserves a definite notion of trajectories for a quantum universe and, with it, a well-defined causal order. It is therefore unlikely that quantum gravity, even in its Bohmian form, will alter our critical analysis of ICO from the standpoint of dBB theory.

This conclusion must also be seen in light of the debate over whether ICOs are real or merely simulated in laboratory-based QS implementations \cite{Procopio2015, Goswami2018, PaunkovicVojinovic2020,Abbott2020,Oreshkov2019,Wechs2023,MacLean2017,Zych2019,VilasiniRenner2024,Ormrod2023,Kristjansson2020,Hamette}. While an approach based on quantum gravity is often presented as the ultimate solution to this debate \cite{Zych2019,HamettePRL}, we do not expect this to be the case here, since the Bohmian representation already involves perfectly well-defined causal trajectories for the 3-metric $g_{ij}(x)$.

At a deeper level, the difficulty is partly linguistic: it concerns the very notion of a localised event, depending on whether it is understood operationally or spatio-temporally. As operational approaches are influenced by the Copenhagen interpretation, the concept of an observer remains poorly defined within a relativistic framework if the observer is itself part of space-time. In this respect, a Bohmian approach may prove superior and unambiguous, although the issue clearly warrants further discussion.

A related open problem concerns Novikov self-consistency on closed timelike curves (CTCs)---as in wormhole time machines. It remains unclear whether such self-consistency is a dynamical consequence of the underlying physics or merely an ad hoc selection of globally consistent histories, and whether dBB theory, with its definite trajectories and guidance equation, can provide an ontological mechanism for this consistency without invoking branching, retrocausal signalling, or an additional consistency postulate \cite{Friedman1990,Morris1988}. The same interpretational difficulty arises for the Lugano process of Baumeler, Feix and Wolf \cite{BaumelerFeixWolf2014}, whose classical, logically consistent causal loop violates a causal inequality without yet admitting an unambiguous spacetime interpretation. Clarifying this point is essential before interpreting causal loops as a genuine physical realisation of indefinite causal order.

A similar problem arises in recent studies of systems `delocalised' in time \cite{Oreshkov2019,Ho2019,Valibouse,Felce2022,Wechs2023}, which, according to the authors of \cite{Ho2019,Wechs2023}, could allow causal inequalities to be violated --- something that would seem to preclude any physical representation in terms of causally ordered trajectories. Since this result appears to conflict with dBB theory, we suggest that the issue has not yet reached sufficient maturity to be resolved unambiguously.

All of this calls for further research into the interpretation of causal inequalities and ICOs in these exotic quantum systems in relation to dBB theory. It suggests that dBB theory --- causal and deterministic by definition --- offers an alternative, or even a solution, to the causal paradoxes discussed in recent years in connection with ICOs. Nevertheless, ICO, much like wave-particle duality, may continue to play a useful role in teaching and engineering quantum mechanics --- provided we do not forget that the primary task of physics is to dispel obscurity.\\

\end{document}